\documentclass[apj]{emulateapj}
\usepackage{epsfig}
\received{May 4, 2009}
\accepted{Dec 8, 2009}

\shorttitle{Dynamical Models for Omega Centauri}
\shortauthors{van der Marel \& Anderson}

\newcommand{\etal}{{et al.~}}
\newcommand{\lta}{\lesssim}
\newcommand{\gta}{\gtrsim}
\newcommand{\kms}{\>{\rm km}\,{\rm s}^{-1}}
\newcommand{\masyr}{\>{\rm mas}\,{\rm yr}^{-1}}
\newcommand{\kpc}{\>{\rm kpc}}
\newcommand{\pc}{\>{\rm pc}}
\newcommand{\Msun}{\>{M_{\odot}}}
\newcommand{\Lsun}{\>{L_{\odot}}}

\newcommand{\vrr}{\overline{v_r^2}}
\newcommand{\vtt}{\overline{v_{\theta}^2}}
\newcommand{\vpp}{\overline{v_{\phi}^2}}
\newcommand{\vnn}{\overline{v_{\rm tan}^2}}
\newcommand{\MBH}{M_{\rm BH}}
\newcommand{\tr}{\tilde \rho}
\newcommand{\tP}{\tilde \Phi}
\newcommand{\omegaCen}{$\omega$ Centauri}
\newcommand{\cL}{{\cal L}}
\newcommand{\vlos}{v_{\rm los}}
\newcommand{\vpmr}{v_{\rm pmr}}
\newcommand{\vpmt}{v_{\rm pmt}}
\newcommand{\asqlos}{\langle v_{\rm los}^2 \rangle}
\newcommand{\asqpmr}{\langle v_{\rm pmr}^2 \rangle}
\newcommand{\asqpmt}{\langle v_{\rm pmt}^2 \rangle}
\newcommand{\slos}{{\bar \sigma}_{\rm los}}
\newcommand{\spmr}{{\bar \sigma}_{\rm pmr}}
\newcommand{\spmt}{{\bar \sigma}_{\rm pmt}}
\newcommand{\spm}{{\bar \sigma}_{\rm pm}}
\newcommand{\Spmr}{{\bar \Sigma}_{\rm pmr}}
\newcommand{\Spmt}{{\bar \Sigma}_{\rm pmt}}
\newcommand{\Spm}{{\bar \Sigma}_{\rm pm}}
\newcommand{\Spmmj}{{\bar \Sigma}_{\rm pmmj}}
\newcommand{\Spmmi}{{\bar \Sigma}_{\rm pmmi}}

\begin{document}

\title{New Limits on an Intermediate Mass Black Hole in Omega
Centauri:\\ II. Dynamical Models\footnote{Based on observations with the
NASA/ESA {\it Hubble Space Telescope}, obtained at the Space Telescope
Science Institute, which is operated by AURA, Inc., under NASA
contract NAS 5-26555.}}

\author{Roeland P.~van der Marel \& Jay Anderson}
\affil{Space Telescope Science Institute, 3700 San Martin Drive, 
       Baltimore, MD 21218}


\begin{abstract}
We present a detailed dynamical analysis of the projected density and
kinematical data available for the globular cluster {\omegaCen}. We
solve the spherical anisotropic Jeans equation for a given density
profile to predict the projected profiles of the RMS velocity ${\bar
\sigma}(R)$, in each of the three orthogonal coordinate directions
(line of sight, proper motion radial, and proper motion tangential).
The models allow for the presence of a central dark mass, such as a
possible intermediate-mass black hole (IMBH). We fit the models to new
HST star count and proper motion data near the cluster center
presented in Paper~I, combined with existing ground-based measurements
at larger radii. The projected density profile is consistent with
being flat near the center, with an upper limit $\gamma \lta 0.07$ on
the central logarithmic slope. The RMS proper motion profile is also
consistent with being flat near the center. The velocity anisotropy
profile, distance, and stellar mass-to-light ratio are all tightly
constrained by the data, and found to be in good agreement with
previous determinations by van de Ven et al. To fit the kinematics we
consider anisotropic models with either a flat core ($\gamma = 0$) or
a shallow cusp ($\gamma = 0.05$). Core models provide a good fit to
the data with $\MBH = 0$; cusp models require a dark mass. If the dark
mass in cusp models is an intermediate-mass black hole (IMBH), then
$\MBH = (8.7 \pm 2.9) \times 10^3 \Msun$; if it is a dark cluster,
then its extent must be $\lta 0.16\pc$. Isotropic models do not fit
the observed proper motion anisotropy and yield spuriously high values
for any central dark mass.  These models do provide a good fit to the
Gauss-Hermite moments of the observed proper motion distributions
($h_4 = -0.023 \pm 0.004$, $h_6 = 0.001 \pm 0.004$). There are no
unusually fast-moving stars observed in the wings of the proper motion
distribution, but we show that this does not strongly constrain the
mass of any possible IMBH. The overall end-result of the modeling is
an upper limit to the mass of any possible IMBH in {\omegaCen}: $\MBH
\lta 1.2 \times 10^4 \Msun$ at $\sim 1\sigma$ confidence (or $\lta 1.8
\times 10^4 \Msun$ at $\sim 3\sigma$ confidence). The $1\sigma$ limit
corresponds to $\MBH/M_{\rm tot} \lta 0.43$\%. We combine this with
results for other clusters and discuss the implications for globular
cluster IMBH demographics. Tighter limits will be needed to rule out
or establish whether globular clusters follow the same black hole
demographics correlations as galaxies. The arguments put forward by
Noyola et al.~to suspect an IMBH in {\omegaCen} are not confirmed by
our study; the value of $\MBH$ they suggested is firmly ruled out.
\end{abstract}




\section{Introduction}
\label{s:intro}

Intermediate-mass black holes (IMBHs) are of interest in a variety of
astrophysical contexts (see, e.g., the reviews by van der Marel 2004;
Miller \& Colbert 2004). Especially the possible presence of IMBHs in
the centers of globular clusters has intrigued astronomers for decades
(e.g., Bahcall \& Wolf 1976). This subject area underwent a
considerable resurgence in recent years due to a combination of two
advances. First, theoretical modeling suggested plausible formation
scenarios for IMBHs in the centers of globular clusters from realistic
initial conditions (Portegies Zwart \& McMillan 2002). Second, the
quality of observational data sets, especially those obtained with the
Hubble Space Telescope (HST), advanced to the point where it is
possible to measure velocity dispersions at a spatial resolution
comparable to the sphere of influence size for plausible IMBH
masses. This led to the identification of three globular clusters with
candidate IMBHs, namely M15, G1, and {\omegaCen}. 

Gerssen \etal (2002) used line-of-sight velocity data for individual
stars in the core-collapsed cluster M15 (NGC 7078) to argue for the
presence of $\sim 3000 \Msun$ in dark material near the center,
possibly in the form of an IMBH. Dull \etal (1997) and Baumgardt \etal
(2003a) showed that this could be a plausible result of
mass-segregation, and need not be an IMBH. Moreover, Baumgardt \etal
(2005) argued that the observed steep brightness profile of M15,
typical of the class of core-collapsed clusters, is in fact not
consistent with expectation for a relaxed distribution of stars around
an IMBH. Ho, Terashima, \& Okajima (2003) and Bash \etal (2008) failed
to detect X-ray or radio emission coincident with the cluster
center. This is not necessarily inconsistent with an IMBH, since the
accretion rate and/or the accretion efficiency are likely to be
low. Nonetheless, M15 does not appear to be as strong an IMBH
candidate as once thought.

Gebhardt \etal (2002, 2005) used the velocity dispersion measured from
integrated light near the center of the M31 cluster G1 to argue for
the presence of a $(1.8 \pm 0.5) \times 10^4 \Msun$ dark mass at the
cluster center. The relaxation time of G1 is too long to have produced
such a concentration of dark mass due to mass segregation. Therefore,
this result was taken as evidence of for the presence of an
IMBH. Baumgardt \etal (2003b) disagreed with this interpretation, and
argued that models with no dark mass (IMBH or otherwise) provided an
equally acceptable fit. The crux of this disagreement lies in the
extremely small size of the sphere of influence ($G \MBH / \sigma^2
\approx 0.03''$) of the putative IMBH, which is less than the spatial
resolution of the observations. Phrased differently, the IMBH induces
only a small signature in the data. As a result, the interpretation of
the data is sensitive to the exact details of the data-model
comparison. The statistical significance of the IMBH detection is not
high, especially if one takes into account the possibility of ``black
hole bias'' (discussed later in Section~\ref{ss:IMBH}). On the other
hand, the possible presence of an IMBH in G1 recently gained further
credence with the detection of weak X-ray and radio emission from near
the cluster center (Pooley \& Rappaport 2006; Kong 2007; Ulvestad,
Greene \& Ho 2007). The emission properties are consistent with
accretion onto an IMBH, but other explanations cannot be ruled
out.

Noyola \etal (2008; hereafter NGB08) argued for the presence of an
IMBH in the center of the cluster {\omegaCen} (NGC 5139), which is
well known for its many enigmatic properties (e.g., Meylan 2002). The
dynamics of this cluster were studied previously by van de Ven
\etal (2006; hereafter vdV06) using sophisticated axisymmetric models. 
However, there was insufficient kinematical data at small radii to
meaningfully constrain the possible presence of any dark mass. NGB08
performed integrated-light measurements of the line-of-sight velocity
dispersion for two $5 \times 5''$ fields, one on the cluster center
that they had determined, and one at $R=14''$ from that center.  They
obtained $\sigma_{\rm los} = 23.0 \pm 2.0 \kms$ and $\sigma_{\rm los}
= 18.6 \pm 1.6 \kms$ for these fields, respectively.  The increase in
$\sigma_{\rm los}$ towards the cluster center was interpreted to imply
the presence of an IMBH of mass $\MBH = 4.0_{-1.0}^{+0.75}
\Msun$. They also determined the surface brightness profile of
{\omegaCen} from HST images, and inferred a shallow central cusp of
logarithmic slope $\gamma = 0.08 \pm 0.03$.  This is consistent with
theoretical predictions for the cusp induced by an IMBH (Baumgardt
\etal 2005).

Both G1 and {\omegaCen} are somewhat atypical globular clusters. They
are the most massive clusters of their parent galaxies, M31 and the
Milky Way, respectively. Both clusters have been suggested to be the
stripped nuclei of larger galaxies (e.g., Freeman 1993; Meylan \etal
2001). Indeed, the nuclei of galaxies have very similar structural
properties to globular clusters (these nuclei are therefore also
referred to as nuclear star clusters; e.g., Walcher \etal
2005). Accretion activity is known to exist in at least some of these
nuclei (e.g., Seth \etal 2008). So if G1 and {\omegaCen} contain
IMBHs, then these may simply be the low-mass tail of the distribution
of super-massive black holes known to exist in galactic nuclei. As
such, they would not necessarily be telling us anything fundamentally
new about the formation and evolution of globular clusters in general.
Nonetheless, G1 and {\omegaCen} are probably the best current
candidates for globular clusters with IMBHs. Validating the existence
of their suggested IMBHs is therefore of fundamental importance.

The (initial) IMBH evidence for M15, G1, and {\omegaCen} was all based
on studies of line-of-sight velocities. Such studies are complicated
because they require spectroscopy in regions of high stellar
density. Integrated-light measurements work well for a cluster as
distant as G1 (Gebhardt \etal 2002, 2005), but the downside is that
the sphere of influence in such clusters is poorly resolved. In closer
Milky Way clusters, the limited number of stars (especially brighter
ones) introduces significant shot noise in integrated-light
measurements. For {\omegaCen}, this required special treatment of the
data to counteract this (NGB08).  For M15, HST/STIS was used to obtain
spectroscopy of individual stars (van der Marel \etal 2002). However,
the signal-to-noise requirements for a velocity measurement limit the
data set to the brightest stars. In turn, this sets the smallest
radius inside which a velocity dispersion can be measured.

A significant improvement in data quality is possible with proper
motion measurements. Such studies do not require much observing time
and allow measurement even of the more numerous faint stars. Moreover,
the ratio of the velocity dispersions in the tangential and radial
proper motion directions directly constrains the velocity-dispersion
anisotropy of the system. This is important because of the well-known
mass-anisotropy degeneracy for stellar systems (Binney \& Mamon
1982). To determine accurate proper motions close to the center
requires an observing platform with high spatial resolution and
long-term stability. The HST provides such a platform. McNamara,
Harrison, \& Anderson (2003) measured proper motions for M15, which
were modeled by van den Bosch \etal (2006) and Chakrabarty
(2006). This confirmed the presence of dark mass near the cluster
center, but did not shed new light on the issue of whether this is due
to mass segregation or an IMBH. McLaughlin \etal (2006) obtained and
modeled proper motions for 47 Tuc, and used these to set an upper
limit of 1000--$1500 \Msun$ on the mass of any IMBH in that cluster.

In Anderson \& van der Marel (2009; hereafter Paper I) we presented
the results of a new observational HST study of {\omegaCen}. For the
central few arcmin of the cluster we determined the projected number
density distribution of some $10^6$ stars, as well as accurate proper
motions of some $10^5$ stars.  The exact position of the cluster
center was determined in three independent ways. First, we determined
the center of the number density distribution using various methods,
taking care to accurately account for incompleteness. Second, we
determined the kinematical center, being the symmetry point of the RMS
proper motion velocity field. And third, we determined the center of
unresolved light seen in 2MASS data. All determinations agree to
within 1--$2''$, yielding a final estimate with an uncertainty of
$\sim 1''$.

The NGB08 center position, and hence the position of their central
spectroscopic field, is at $R=12''$ from the center determined in
Paper~I (and coincidentally, the second off-center spectroscopic field
they studied is also at $R=12''$). The differences between the centers
determined in these two studies is likely due to a combination of
subtle effects that biased the NGB08 analysis, as discussed in
Paper~I. As a result of this, the density cusp found by NGB08 was not
measured around the actual cluster center. Also, we showed in Paper~I
that their use of unresolved light (as opposed to star counts)
increased the impact of shot noise. In Paper~I we determined the
number density profile around the actual center of {\omegaCen}. It was
found to be well matched by a standard King model, even though a
shallow cusp may not be ruled out.

In Paper~I we also determined the RMS proper motions of stars detected
in the same $5'' \times 5''$ fields studied spectroscopically by
NGB08. For the canonical distance $D=4.8 \kpc$ (vdV06), the results
translate to one-dimensional values $\spm = 18.9
\pm 1.3 \kms$ and $\spm = 19.0 \pm 1.6 \kms$, for their
``central'' and ``off-center'' fields respectively. So we did not
confirm the NGB08 finding that one of their fields has a higher
dispersion. The observation that there is no significant difference in
proper motion dispersion between the two fields is consistent with our
finding that they are both at the same radius $R=12''$ from the actual
cluster center.

Our results of Paper~I did not confirm the arguments put forward by
NGB08 to suspect the presence of an IMBH in {\omegaCen}. However, this
does not mean that such an IMBH may not be present after all. The size
and quality of our proper motion data set are superb. This provides
the opportunity to study the central dynamics of {\omegaCen} at a
level of detail that is unmatched by other clusters. We therefore
present here a detailed study of the dynamics of {\omegaCen}, with the
primary goal of exploiting the new data from Paper~I to constrain the
mass of any possible IMBH. We also include existing ground-based data
in our data-model comparisons, to obtain the best possible results.
   
We restrict our modeling efforts in the present paper to spherical
models. This is a simplification, given that {\omegaCen} is elongated
and rotating (in fact, unusually so for a globular cluster). However,
both the elongation (Geyer, Nelles, \& Hopp 1983) and the rotation
rate (vdV06) decrease towards the cluster center. Therefore, the
assumption of sphericity is likely to be reasonable in the central
region most relevant for constraining an IMBH. We find support for
this throughout our paper by showing that spherical models yield an
anisotropy profile, distance, and mass-to-light ratio that are in
excellent agreement with the values obtained by vdV06. More
sophisticated models, such as those of the type described by vdV06,
van den Bosch \etal (2006) and Chaname \etal (2008), have the
potential to constrain other aspects of the structure of {\omegaCen},
such as its inclination, rotation properties, and the presence of any
kinematic subcomponents. However, spherical models are sufficient to
obtain robust constraints on the central mass distribution.

The structure of this paper is as follows. Section~\ref{s:modeling}
discusses the general modeling approach. Section~\ref{s:sbdata}
presents the data on star count and surface brightness profiles used
for the analysis. Section~\ref{s:sbmodel} discusses parameterized fits
to these data. Section~\ref{s:kinematics} presents the kinematical
data used for the analysis. Section~\ref{s:compare} presents the
kinematical data-model comparison, with a determination of the
best-fitting model parameters. Section~\ref{s:demographics} discusses
how the new IMBH results for {\omegaCen} fit in with our general
understanding of IMBH demographics in globular clusters.
Section~\ref{s:conc} presents and discusses the conclusions.

\section{Modeling Approach}
\label{s:modeling}

\subsection{Mass Density}
\label{ss:dens}

Imaging observations of a cluster yield an estimate of $\mu_{{\rm
obs,}X}(R)$, the radial profile of the surface brightness in a
photometric band $X$, averaged over concentric circles of radius $R$
around the cluster center. This quantity is related to the projected
intensity $I_{\rm obs}(R)$ according to
\begin{equation}
\label{obsmag}
  \mu_X [{\rm mag} \> {\rm arcsec}^{-2}] = 
      -2.5 \log I_{\rm obs} [L_{\odot} \> {\rm pc}^{-2}] 
      + 21.572 + M_{\sun, {\rm X}} ,
\end{equation}
where $M_{\sun,X}$ is the absolute magnitude of the sun in the given
photometric band. Throughout this paper, units in which quantities are
expressed are given in square brackets where relevant. In the presence
of $A_X$ magnitudes of foreground extinction, the extinction-corrected
profile is given by
\begin{equation}
\label{projint}
  I(R) = 10^{0.4 A_{\rm X}} \> I_{\rm obs}(R)  .
\end{equation}

\null
For a spherical cluster, $I(R)$ is the line-of-sight integration over
the three-dimensional luminosity density $j(r)$. This integration can
be inverted through an Abel transform (e.g., Binney \& Tremaine 1987)
to yield
\begin{equation}
\label{abel}
   j(r) = -{1\over\pi} \int_{r}^{\infty} \left [ { {dI}\over{dR} }
          \right ] (R) { {dR} \over {\sqrt{R^2-r^2}} } .
\end{equation}
To express this quantity in units of $L_{\odot} \> {\rm pc}^{-3}$ one
needs to specify the cluster distance $D$. Sizes in arcsec and pc are
related according to
\begin{equation}
\label{radscale}
   R[{\rm pc}] = R[{\rm arcsec}] \> \pi \> D[{\rm kpc}] / 648 .
\end{equation}

\null
The mass density of stellar objects and their remnants is 
\begin{equation} 
  \rho(r) = [M/L](r) \> j(r) ,
\end{equation}
where $[M/L](r)$ is the average mass-to-light ratio at each
radius. Gradients in $[M/L](r)$ can be thought of as arising from two
separate ingredients: (i) mass segregation among luminous stars; and
(ii) mass segregation of the heavy compact stellar remnants (heavy
white dwarfs, neutron stars, and stellar-mass black holes) with
respect to the luminous stars. The first ingredient generally causes a
shallow gradient in $[M/L](r)$ over the scale of the entire cluster
(e.g., Gill \etal 2008). This has little effect on the model
predictions in the area surrounding the very center, where we wish to
search for the signature of an IMBH. By contrast, the second
ingredient can in some clusters yield a strong increase in $[M/L](r)$
near the very center (e.g., Dull \etal 1997; Baumgardt
\etal 2003a) that mimics the presence of a dark sub-cluster. 

Our modeling approach allows for exploration of models with arbitrary
$[M/L](r)$. However, we have found it useful and sufficient in the
present context to restrict attention to a parameterized subset. Based
on the preceding considerations, we write the mass density as
\begin{equation} 
  \rho(r) \equiv \tr(r) + \rho_{\rm dark} , \qquad 
  \tr(r) \equiv \Upsilon j(r) .
\end{equation}
Here $\tr(r)$ is the mass density obtained under the assumption of a
constant mass-to-light ratio $\Upsilon$. The excess $\rho_{\rm dark}$
is treated mathematically as a separate component. For $\rho_{\rm
dark}$ we have found it convenient to use the parameterization
\begin{equation}
\label{rhodark}
  \rho_{\rm dark} = ( 3 M_{\rm dark} a_{\rm dark}^2 / 4 \pi ) /
                     (r^2 + a_{\rm dark}^2)^{5/2} ,
\end{equation}
corresponding to a Plummer density distribution. Here $M_{\rm dark}$
is the total excess mass in dark objects and $a_{\rm dark}$ is the
characteristic scale length of their spatial distribution. The
quantities $\Upsilon$, $M_{\rm dark}$ and $a_{\rm dark}$ provide a
convenient way for exploring a range of plausible $[M/L](r)$ functions
using only three parameters.

By assuming that $\Upsilon$ is independent of $r$, we neglect mass
segregation among the luminous stars. For {\omegaCen} there are in
fact several pieces of evidence that it has not yet undergone complete
mass segregation, consistent with its long half-mass relaxation time
of $\sim 10^{9.96 \pm 0.03}$ years (McLaughlin \& van der Marel
2005). First, Merritt, Meylan \& Mayor (1997) constructed
non-parametric dynamical models for {\omegaCen}. They found that at
the large radii sampled by their data, $\rho(r) \propto j(r)$ to
within the error bars. Second, vdV06 constructed more general
dynamical models for a wider range of data, and they too found that at
all radii sampled by their study $[M/L](r)$ is consistent with being
constant. Third, Anderson (2002) studied luminosity functions (LFs) at
different radii in {\omegaCen} using data obtained with HST. He found
that variations in the LFs are less than would be expected if mass
segregation had proceeded to the point of energy equipartition. And
fourth, in Paper I we studied the proper motion dispersion of stars
along the main sequence. This showed that lower mass stars do move
faster, but by a smaller amount than would be expected for energy
equipartition.

\subsection{Gravitational Potential}
\label{ss:pot}

The total gravitational potential of the system is
\begin{equation}
  \Phi_{\rm tot} = \tP(r) + \Phi_{\rm dark}(r) - G \MBH / r  , 
\end{equation}
where $G$ is the gravitational constant and $\MBH$ is the mass of any
central IMBH. The gravitational potential corresponding to $\tr(r)$ is
\begin{equation}
   \tP(r) = -4 \pi G \left [ {1\over r} \int_{0}^{r}
      \tr(r') r'^2 \> dr' + \int_{r}^{\infty} \tr(r') r' \>
      dr' \right ] .
\end{equation}
The Plummer potential corresponding to $\rho_{\rm dark}$ is 
\begin{equation}
   \Phi_{\rm dark}(r) = - G M_{\rm dark} / (r^2 + a_{\rm dark}^2)^{1/2} .
\end{equation}

\null The potential $\Phi_{\rm tot}$ could in principle be expanded to
also include an extended halo of particulate dark matter, but that is
more relevant for galaxies. There is no evidence that star clusters
are embedded in dark halos. For the specific case of {\omegaCen},
vdV06 showed that there is no need for dark matter out to at least two
half-light radii.

\subsection{Intrinsic Velocity Moments}
\label{ss:dyn}

The Jeans equation for hydrostatic equilibrium in a spherical stellar
system is (e.g., Binney \& Tremaine 1987)
\begin{equation}
\label{jeans}
  {{d(\nu \vrr)}\over{dr}} + 2 {{\beta \nu \vrr} \over {r}} = -\nu
    {{d\Phi_{\rm tot}}\over{dr}} ,
\end{equation}
where $\nu(r)$ is the number density of the stars under
consideration. Because of the spherical symmetry, $\vtt = \vpp$. The
function $\beta(r) \equiv 1 - \vnn/\vrr$ measures the anisotropy in
the second velocity moments, where $\vnn \equiv \vtt =
\vpp$. Models with $\beta =0$ are isotropic, models with $\beta < 0$
are tangentially anisotropic and models with $0 < \beta \leq 1$ are
radially anisotropic. The models may in principle be rotating if
$\overline{v_{\phi}} \not= 0$. However, for the purposes of the
present paper we do not need to address how the second azimuthal
velocity moment $\vpp \equiv \sigma_{\phi}^2 + \overline{v_{\phi}}^2$
splits into separate contributions from random motions and mean
streaming.

The measured kinematics in a cluster are generally representative for
only a subset of the stars (e.g., stars in that magnitude range for
which kinematics can be measured). The number density of this subset
is
\begin{equation}
\label{nuj}
   \nu(r) = j(r) f(r) / \lambda(r) ,
\end{equation}
where $\lambda(r)$ is the average luminosity of a star in the subset
and $f(r)$ is the fraction of the total light that the subset
contributes at each radius. Compact stellar remnants play no role in
this equation, since they contribute neither to $\nu(r)$ (their
kinematics are not probed) nor to $j(r)$ (they emit negligible amounts
of light). However, $f(r)$ and $\lambda(r)$ can vary slowly as a
function of radius due to mass segregation among the luminous
stars. The quantity $\lambda(r)$ may also be affected by observational
selection effects that vary as a function of radius. Nonetheless,
variations in $\lambda(r)$ will generally be smaller than those in
$f(r)$, as long as kinematics are measured for stars of similar
brightness at each radius. Both $f(r)$ and $\lambda(r)$ are only
expected to have a shallow gradient over the scale of the entire
cluster. These gradients are much smaller than those in $j(r)$, which
can vary by orders of magnitude. Moreover, in the present paper we are
mostly concerned with the small area surrounding the very center. We
therefore assume that
\begin{equation}
\label{flam}
  f(r) / \lambda(r) = {\rm constant} , 
\end{equation}
which implies that, as before, we neglect mass segregation among the
luminous stars. 

It follows from the preceding that one can substitute $j(r)$ for
$\nu(r)$ in the Jeans equation~(\ref{jeans}), without actually needing
to know the value of the constant in equation~(\ref{flam}). The Jeans
equation is then a linear, first-order differential equation for $j
\vrr$ with solution
\begin{equation}
\label{sigrdef}
  j \vrr (r) = \int_{r}^{\infty} j(r') 
              \Bigl [ {{d\Phi_{\rm tot}}\over{dr}} \Bigr ](r') \>
         \exp \Bigl [ 2 \int_{r}^{r'} {{\beta(r'') \> dr''}\over{r''}} \Bigr ]
         \> dr'.
\end{equation}
This is a special case of a more general solution for certain classes of
axisymmetric systems (Bacon, Simien \& Monnet 1983). We consider anisotropy
functions of the form
\begin{equation}
\label{betadef}
  \beta(r) = (\beta_{\infty} r^2 + \beta_0 r_a^2) / (r^2 + r_a^2) , 
\end{equation}
where $r_a >0$ is a characteristic anisotropy radius, $\beta_0 \leq 1$
is the central value of $\beta(r)$, and $\beta_{\infty} \leq 1$ is the
asymptotic value at large radii. If $\beta_0 = \beta_{\infty}$ then
$\beta(r)$ is constant. With this $\beta(r)$, equation~(\ref{sigrdef})
reduces to
\begin{eqnarray}
\label{jeanssol}
  j \vrr & & (r) = \int_{r}^{\infty} j(r')
       \left [ {{d\Phi_{\rm tot}}\over{dr}} \right ](r') \> \times \nonumber \\
  & &\left [ {{r'^2 + r_a^2} \over {r^2 + r_a^2}} \right ]^{\beta_{\infty}}
     \left [ \Bigl ( {{r'^2}\over{r^2}} \Bigr ) \Big /
             \Bigl ({{r'^2 + r_a^2} \over {r^2 + r_a^2}} \Bigr )
     \right ]^{\beta_0} \> dr'  , 
\end{eqnarray}
which is always positive. 

The integral in equation~(\ref{jeanssol}) can be used with the
equations given for $j(r)$ and $\Phi_{\rm tot}(r)$ in
Sections~\ref{ss:dens} and~\ref{ss:pot} to allow numerical evaluation
of $\vrr(r)$. This requires knowledge of the observed surface
brightness profile $\mu_{{\rm obs,}X}(R)$ and the known extinction
$A_X$. It also requires choices for the scalar model quantities $D$,
$\Upsilon$, $\MBH$, $M_{\rm dark}$, $a_{\rm dark}$, $\beta_0$,
$\beta_{\infty}$ and $r_a$, which can all be chosen to optimize the
fit to the available kinematical data.

\subsection{Projected Velocity Moments}
\label{ss:proj}

To calculate quantities that are projected along the line of sight,
consider an arbitrary point P in the cluster. We adopt a Cartesian
coordinate system (pmr, los, pmt) with three orthogonal axis and its
origin at the cluster center C. The pmr-axis (proper motion radial)
lies in the plane of the sky and points from the projected position of
C to the projected position of P. The los-axis (line of sight) points
from the observer to the point C. The pmt-axis (proper motion
tangential) lies in the plane of the sky, and is orthogonal to the
pmr- and los-axes in a right-handed sense. The cylindrical polar
coordinate system associated with (pmr, los, pmt) is $(R,\phi,z)$ and
the spherical polar coordinate system is $(r,\theta,\phi)$, with
$\theta \in [-\pi/2,\pi/2]$, and $\theta = \pi/2$ corresponding to the
pmt axis. The velocities in the (pmr, los, pmt) system satisfy
\begin{eqnarray}
  \vpmr & = & v_r \cos \phi - v_{\phi} \sin \phi , \nonumber \\
  \vlos & = & v_r \sin \phi + v_{\phi} \cos \phi , \nonumber \\
  \vpmt & = & v_{\theta} .
\end{eqnarray}
Since $\overline{v_r v_{\phi}} = 0$ in a spherical system, the
second moments are
\begin{eqnarray}
  \overline{\vpmr^2} (r,\phi) & = & 
          \vrr(r) [1 - \beta(r) \sin^2 \phi] , \nonumber \\
  \overline{\vlos^2} (r,\phi) & = & 
          \vrr(r) [1 - \beta(r) \cos^2 \phi] , \nonumber \\
  \overline{\vpmt^2} (r,\phi) & = & 
          \vrr(r) [1 - \beta(r)]  ,
\end{eqnarray}
where we have used that $\vpp = \vtt = \vrr [1 - \beta(r)]$. 

The projected second moments satisfy
\begin{eqnarray}
\label{pmom}
  I \asqpmr (R) & = & \int_R^{\infty} 
        {{2 r j \vrr (r) [1 - \beta(r) + \beta(r) (R/r)^2) ] \> dr}\over
        {\sqrt{r^2-R^2}}} , \nonumber \\
  I \asqlos (R) & = & \int_R^{\infty} 
        {{2 r j \vrr (r) [1 - \beta(r) (R/r)^2 ] \> dr}\over
        {\sqrt{r^2-R^2}}} , \nonumber \\
  I \asqpmt (R) & = & \int_R^{\infty} 
        {{2 r j \vrr (r) [1 - \beta(r)] \> dr}\over
        {\sqrt{r^2-R^2}}} .
\end{eqnarray}
The derivations of these equations use the facts that $\cos^2 \phi =
(R/r)^2$ and $d\>{\rm los} = d \sqrt{r^2-R^2} = r \> dr
/\sqrt{r^2-R^2}$. The factor of two arises from the identical
contributions from points with los $<0$ and los $>0$. The angle
brackets around the squared velocities on the left-hand side indicate
the combined effect of averaging over velocity space and
density-weighted averaging along the line of sight. Formally speaking,
the weighting is with number density if one is modeling discrete
measurements, and with luminosity density if one is modeling
integrated light measurements.  However, in the present paper we are
assuming that $\nu(r) \propto j(r)$, so the two cases are equivalent.
The integrals in equation~(\ref{pmom}) can be evaluated numerically
once the Jeans equation has been solved.

\subsection{Projected Velocity Distributions}
\label{ss:vp}

The normalized projected velocity distributions $\cL(v,R)$ at
projected radius $R$ are generally different for the los, pmr, and pmt
directions. The Jeans equation suffices to calculate the second
velocity moments of these distributions. To calculate the
distributions themselves it is necessary to know the full phase-space
distribution function. This is not generally straightforward (or
unique) for anisotropic models. However, for an isotropic model the
distribution function depends only on the energy, and it can be
uniquely calculated for given $j(r)$ (being proportional to both
$\nu(r)$ and $\tr(r)$ under the present assumptions) and $\Phi_{\rm
tot}(r)$, with the help of Eddington's equation (Binney \& Tremaine
1987). Because of the isotropy, the distribution $\cL_{\rm iso}(v,R)$
is independent of the choice of the velocity direction (los, pmr, or
pmt) and satisfies, in analogy with equation~(\ref{pmom}),
\begin{equation}
\label{VP}
 I(R) \cL_{\rm iso}(v,R) = 
      \int_R^{r_{\rm max}(v)} 
        {{2 r j(r) \cL_{\rm loc}(v,r) \> dr}\over{\sqrt{r^2-R^2}}} .
\end{equation}
Here $r_{\rm max}(v)$ is defined to satisfy $\Psi(r_{\rm max})
\equiv {1\over2}v^2$, with $\Psi \equiv -\Phi_{\rm tot}$. The quantity 
$\cL_{\rm loc}(v,r)$ is the local velocity distribution (before
projection along the line of sight) at radius $r$, which can be
written with the help of Eddington's equation as (van der Marel 1994a)
\begin{equation}
\label{locVP}
   j(r) \cL_{\rm loc}(v,r) = {1\over{\pi\sqrt{2}}} 
          \int_0^{\Psi(r)-{1\over2}v^2} 
          \Bigl [ {{d j}\over{d\Psi}} \Bigr ]_{\Psi'}
          {{d\Psi'}\over{\sqrt{\Psi(r)-{1\over2}v^2-\Psi'}}}.
\end{equation}

\null
The integrals in equations~(\ref{VP}) and~(\ref{locVP}) can be used
with the equations given for $j(r)$ and $\Phi_{\rm tot}(r)$ in
Sections~\ref{ss:dens} and~\ref{ss:pot} to allow numerical evaluation
of $\cL_{\rm iso}(v,R)$. The result is a symmetric function of $v$
[i.e., $\cL_{\rm iso}(v,R) = \cL_{\rm iso}(-v,R)$]. This remains true
if the distribution function is given a non-zero odd part in the
angular momentum $L_z$ around some axis (corresponding to a rotating
model), provided that $\cL_{\rm iso}(v,R)$ is then interpreted as the
distribution averaged over a circle of radius $R$.

\subsection{Observed quantities}
\label{ss:obsquant}

Equations~(\ref{pmom}) and~(\ref{VP}) define velocity moments and
velocity distributions at a fixed position on the projected plane of
the sky. However, to determine these quantities observationally it is
necessary to gather observations over a finite area. In
integrated-light measurements this happens naturally through the use
of, e.g., apertures, fibers, or pixels along a long slit of finite
width. In discrete measurements of individual stars this happens by
binning the data during analysis. In either case, the corresponding
model predictions are obtained as a density-weighted integral over the
corresponding area in the projected $(x,y)$ plane of the sky. For
example, the second line-of-sight velocity moment over an aperture is
\begin{equation}
\label{aperav}
  \asqlos_{\rm ap} = \int_{\rm ap} I\asqlos(R) \> dx \> dy \Big / 
                     \int_{\rm ap} I(R)        \> dx \> dy  .
\end{equation}
Similar equations apply to the other velocity moments and to the
velocity distribution $\cL_{\rm iso}(v,R)$. In integrated light
measurements one can also take the observational point-spread-function
(PSF) into account by convolving both $I(x,y)$ and $I\asqlos(x,y)$
with the PSF before evaluation of the integrals in
equation~(\ref{aperav}).

In the following we write 
\begin{equation}
  \slos \equiv \asqlos_{\rm ap}^{1/2} , \quad
  \spmr \equiv \asqpmr_{\rm ap}^{1/2} , \quad
  \spmt \equiv \asqpmt_{\rm ap}^{1/2} .
\end{equation}
It is also useful to consider the quantity
\begin{equation}
\label{pmtot}
  \spm \equiv \langle ( v_{\rm pmr}^2 + v_{\rm pmt}^2 ) / 2 
              \rangle_{\rm ap}^{1/2}
       = \sqrt{(\spmr^2 + \spmt^2)/2} ,
\end{equation}
which corresponds to the one-dimensional RMS velocity averaged over
all directions in the plane of the sky. Note that velocities in the
plane of the sky are not directly accessible observationally. Instead
of $\spmr$ and $\spmt$ we measure proper motions $\Spmr$ and $\Spmt$
that satisfy
\begin{equation}
\label{pmunit}
  {\bar \Sigma}_{\ldots} [\masyr] = {\bar \sigma}_{\ldots} [\kms] \> /
  \> ( 4.7404 \> D[\kpc] ) .
\end{equation}
Transformation to velocities requires an assumption about the cluster
distance $D$.

The overlying bar in the preceding equations is used to differentiate
the root-mean-square (RMS) velocity ${\bar \sigma}$ from the velocity
dispersion $\sigma$. These quantities are equal only if the mean
streaming is zero (which is always true in the pmr direction).

\subsection{Optimizing and Assessing the Fit to the Data}
\label{ss:scaling}

In general, we will want to model line-of-sight datapoints ${\bar
\sigma}_{{\rm los},i} \pm \Delta {\bar \sigma}_{{\rm los},i}$ available for
$i=1,\ldots,N$, as well as proper motion datapoints ${\bar
\Sigma}_{{\rm pm\ldots},j} \pm \Delta {\bar \Sigma}_{{\rm pm\ldots},j}$ 
available for $j=1,\ldots,M$ (with each proper motion data point
referring to either the pmr or the pmt direction). To calculate model
predictions we use the known surface brightness profile $\mu_{{\rm
obs},X}(R)$ and the extinction $A_X$. The questions are then how to
assess the fit to the data for given model parameters, and how to
determine the parameters that provide adequate fits.

\subsubsection{Distance and Mass-to-Light Ratio}
\label{sss:distups}

Suppose that we have obtained model predictions ${\hat \sigma}_{{\rm
los},i}$ and ${\hat \Sigma}_{{\rm pm\ldots},j}$ for the datapoints,
using trial values ${\hat D}$ and ${\hat \Upsilon}$ for the distance
and mass-to-light ratio, respectively. The fit can then be improved by
scaling of the model predictions, which corresponds to rescaling of
the mass and distance scales of the model. Let $F_{\rm los}$ and
$F_{\rm pm}$ be factors that scale the model predictions in the
line-of-sight and proper motion directions, respectively. The
best-fitting values of these scale factors and their error bars
can be calculated using the statistics
\begin{eqnarray}
\label{chisqlospm}
  \chi^2_{\rm los} & \equiv & \sum_i 
     [ ({\bar \sigma}_{{\rm los},i} - 
            F_{\rm los} {\hat \sigma}_{{\rm los},i} ) /
       \Delta {\bar \sigma}_{{\rm los},i} ]^2 , \nonumber \\
  \chi^2_{\rm pm} & \equiv & \sum_j
     [ ({\bar \Sigma}_{{\rm pm\ldots},j} - F_{\rm pm} 
        {\hat \Sigma}_{{\rm pm\ldots},j} ) /
       \Delta {\bar \Sigma}_{{\rm pm\ldots},j} ]^2 .
\end{eqnarray}
The best fit in each case is the value with the minimum $\chi^2$
(i.e., $d\chi^2/dF_{\ldots} = 0$), and the 68.3\% confidence region
corresponds to $\Delta \chi^2 = \chi^2 - \chi^2_{\rm min} \leq 1$. The
inferred scalings imply that the distance and mass-to-light ratio are
optimized for $D_{\rm fit} = {\hat D} F_{\rm los} / F_{\rm pm}$ and
$\Upsilon_{\rm fit} = {\hat \Upsilon} F_{\rm los} F_{\rm pm}$. These
fits apply at the given values of the other model parameters, some of
which are also affected by the scaling. The model parameters $\MBH$
and $M_{\rm dark}$ are proportional to $F_{\rm los}^3 / F_{\rm pm}$,
and the model parameter $a_{\rm dark}$ is proportional to $F_{\rm los}
/ F_{\rm pm}$. 

The error bars in $D_{\rm fit}$ and $\Upsilon_{\rm fit}$ (at fixed
values of the other parameters) follow from propagation of the
uncertainties in $F_{\rm los}$ and $F_{\rm pm}$. The errors in $F_{\rm
los}$ and $F_{\rm pm}$ are uncorrelated, but the errors in $D_{\rm
fit}$ and ${\hat \Upsilon}$ can be highly correlated or
anti-correlated. We will not generally quote the correlation
coefficient in this paper, but nonetheless, this should be kept in
mind where necessary.

The use of scalings as described here provides a means of reducing the
parameter space of the models by two. Also, it directly yields the
best-fit distance and mass-to-light ratio if the remaining parameters
can be determined or eliminated independently. The parameters $\MBH$,
$M_{\rm dark}$ and $a_{\rm dark}$ can be determined as described in
Section~\ref{sss:dark}. They primarily affect the datapoints near the
cluster center. So they can also be effectively eliminated by
restricting the definition of the $\chi^2$ quantities in
equation~(\ref{chisqlospm}) to the datapoints at large radius. The
anisotropy profile $\beta(r)$ can be determined as described in
Section~\ref{sss:beta}.

\subsubsection{Dark Mass}
\label{sss:dark}

Any dark component in our models can be either a cluster with
parameters $(M_{\rm dark},a_{\rm dark})$ or an IMBH of mass $\MBH$.
The latter is mathematically (but not astrophysically) equivalent to a
cluster with parameters $(M_{\rm dark} = \MBH , a_{\rm dark} = 0)$.
It is therefore not possible to discriminate between an IMBH and a
dark cluster of sufficiently small size (within a resolution
determined by the properties of the data). Conversely, we only need to
explore the $(M_{\rm dark},a_{\rm dark})$ parameter space, and this
will automatically tell us (at $a_{\rm dark} = 0$) what the limits on
an IMBH are. In the following we will use the term ``dark mass'' to
refer generically to a dark component that could be either a dark
cluster as in equation~(\ref{rhodark}) or an IMBH. We have not
explored models that have both a dark cluster {\it and} an IMBH,
although that would not have been difficult.

At given $\beta(r)$, we start by calculating initial guesses for
${\hat D}$ and ${\hat \Upsilon}$ by fitting only the data at large
radii. We then calculate for a range of trial values $({\hat M}_{\rm
dark},{\hat a}_{\rm dark})$ the model predictions ${\hat \sigma}_{{\rm
los},i}$ and ${\hat \Sigma}_{{\rm pm\ldots},j}$ for all the
datapoints. For each pair of trial values we calculate $D_{\rm fit}$
and $\Upsilon_{\rm fit}$ as in Section~\ref{sss:distups} by using the
quantities $\chi^2_{\rm los}$ and $\chi^2_{\rm pm}$, including
observations at all distances from the cluster center. We then map the
contours of $\chi^2 \equiv \chi^2_{\rm los} + \chi^2_{\rm pm}$ in the
two-dimensional $(M_{\rm dark},a_{\rm dark})$ parameter space. The
best fit $(M_{\rm dark, fit},a_{\rm dark, fit})$ is obtained as the
point of minimum $\chi^2$. The uncertainties of the fit follow from
standard $\Delta \chi^2$ statistics, with the 68.3\% confidence region
corresponding to $\Delta \chi^2 = \chi^2 - \chi^2_{\rm min} \leq 2.3$
(Press \etal 1992). If there are combinations with $M_{\rm dark} = 0$
within the confidence region, then a dark mass is not required to fit
the data, for the assumed $\beta(r)$. If there are combinations with
$M_{\rm dark} \not= 0$ and $a_{\rm dark} = 0$ within the confidence
region, then the presence of an IMBH is consistent with the data, for
the assumed $\beta(r)$. If we assume that the dark mass is an IMBH,
then its 68.3\% confidence region is given by $\Delta \chi^2 \leq 1$
at $a_{\rm dark} = 0$.

The presence of an IMBH not only increases the RMS velocity towards
the center, but it also changes the shape of the velocity
distribution. The distribution measured through an aperture around the
center will have power-law wings that are significantly broader than
those of a Gaussian distribution (van der Marel 1994b). This leads to
the possibility of detecting very high velocity stars (Drukier \&
Bailyn 2003). It is therefore useful to determine and analyze the
observed velocity distribution of the stars around the center, to
further assess the plausibility of the presence of an IMBH. The
observed distribution can be compared to the predicted distribution
$\cL_{\rm iso, ap}(v)$ that can be calculated for an isotropic
model. It is also useful to calculate the Gauss-Hermite moments (van
der Marel \& Franx 1993) of the observed velocity distribution as a
function of distance from the center. A sufficiently massive IMBH
causes an increase in $h_4$ towards the center (van der Marel
1994b). The absence of broad wings or a central increase in $h_4$ can
be used to obtain an upper limit to the mass of any IMBH.

\subsubsection{Anisotropy Profile}
\label{sss:beta}

Binney \& Mamon (1982) showed that there are many different
combinations of $\beta(r)$ and the enclosed mass $M(<r)$ in a
spherical system that can produce the same observed projected RMS
line-of-sight velocity profile $\slos(R)$. The key to constraining the
mass distribution, and hence the possible presence of an IMBH, is
therefore to constrain the anisotropy profile $\beta(r)$. This is not
possible when only RMS line-of-sight velocities are
available. However, $\beta(r)$ can be constrained when information is
available on deviations of the line-of-sight velocity distributions
(LOSVDs) from a Gaussian, as function of projected position on the sky
(van der Marel \& Franx 1993; Gerhard 1993). This information is
fairly indirect, and to exploit it one must construct complicated
numerical models that retrieve the full phase-space distribution
function of the stars (van der Marel \etal 1998; Cretton \etal 1999;
Gebhardt \etal 2000; Valluri \etal 2004; Thomas \etal 2004; van den
Bosch \etal 2008). This has been explored with considerable success in
the context of studies of early-type galaxies (e.g., Cappellari \etal
2007).

In the present context, high-quality proper motion data are available.
It is then considerably more straightforward to constrain
$\beta(r)$. This was demonstrated by Leonard \& Merritt (1989), who
were the first to use proper motion data to examine in detail the mass
distribution in a star cluster. The function $[\Spmt/\Spmr] (R)$ is a
direct measure of anisotropy as seen in the plane of the sky. This
ratio is independent of either the distance or the mass-to-light ratio
scaling of the model. There is only a weak dependence on the presence
of a dark mass (and this only affects the very central
region). Therefore, the one-dimensional observed function
$[\Spmt/\Spmr] (R)$ tightly constrains the one-dimensional intrinsic
function $\beta(r)$. In fact, for a spherical system there is a
unique relation between the two (Leonard \& Merritt 1989). This is the
primary reason why in the present context it is possible to obtain
robust results based on something as simple as the spherical Jeans
equation, while much more sophisticated techniques have become
standard practice for galaxies.

To build some intuitive understanding of the meaning of $[\Spmt/\Spmr]
(R)$, consider the case in which $\beta(r)$ is independent of radius
and $j \vrr(r) \propto r^{-\xi}$. The relevant integrals in
equation~(\ref{pmom}) can then be expressed analytically in terms of
beta-functions. Their ratio is independent of radius $R$ and
simplifies to
\begin{equation}
  [\Spmt/\Spmr] = \xi(1-\beta) / (\xi-\beta) . 
\end{equation}
For small $\beta$ (i.e., a system not too far from isotropy) we can
use a first order Taylor approximation, which yields
\begin{equation}
  \beta_{\rm pm} (R) = \beta (1 - {1\over\xi})  ,
\end{equation}
where we have defined the function 
\begin{equation}
  \beta_{\rm pm} (R) \equiv 1 - \lbrace [\Spmt/\Spmr](R) \rbrace^2 , 
\end{equation}
in analogy with the definition of $\beta (r)$. This implies that the
deviation of $[\Spmt/\Spmr]$ from unity is a factor $(1 -
{1\over\xi})$ of the deviation of $[\vnn/\vrr]^{1/2}$ from
unity. (This uses the Taylor approximation $(1-\beta)^{1/2} \approx 1
- (\beta/2)$, and similarly for $\beta_{\rm pm}$.) Hence,
$[\Spmt/\Spmr(R)]$ is a ``diluted'' measure of the anisotropy that is
present intrinsically, with dilution factor $(1 - {1\over\xi})$.

The approximations made above cannot be used in the core of a cluster,
where $\xi = 0$. The assumption that the system remains scale free all
along the line of sight is then not accurate. However, in the outer
parts of a cluster the approximation is a reasonably accurate guide,
with the value of $\xi$ depending on the exact cluster structure. For
example, an isothermal sphere has $\xi=2$ at large radii, while a
Plummer model has $\xi=6$. The corresponding dilution factors $(1 -
{1\over\xi})$ are $0.50$ and $0.83$, respectively. Real clusters
typically fall between these extremes. More extensive analytical
calculations of $\beta_{\rm pm} (R)$ are presented in, e.g., Leonard
\& Merritt (1989) and Wybo \& Dejonghe (1995). Either way, analytical
calculations are mostly useful here for illustration. In practice we
calculate the integrals in equation~(\ref{pmom}) numerically.

The extent to which the velocity distribution of the stars (in either
the los, pmr, or pmt direction) differs from a Gaussian provides
secondary information on the amount of velocity anisotropy in the
system. The modeling technique used here is insufficient to exploit
this information, since we are not calculating actual phase-space
distribution functions for anisotropic models (although this would in
principle be possible, e.g., Dejonghe 1987; Gerhard 1993; van der
Marel \etal 2000). Nonetheless, we can observationally characterize
the Gauss-Hermite moments as function of projected distance from the
cluster center. Also, we can calculate the predicted velocity
distributions $\cL_{\rm iso}(v,R)$ for an isotropic model, and from
this the corresponding Gauss-Hermite moments. Comparison of the
observed and predicted quantities provides an independent assessment
of possible deviations from isotropy.

\subsection{Numerical implementation}
\label{ss:software}

To apply the formalism described above we augmented a software
implementation used previously in, e.g., van der Marel (1994a). The
code uses a logarithmically spaced grid, from very small to very large
radii (so that a negligible fraction of the cluster mass resides
outside the grid). Relevant quantities are tabulated on this grid.
Integrals are evaluated through Romberg quadrature (e.g., Press \etal
1992) to high numerical precision. Tabulated quantities are
interpolated using cubic splines for use in subsequent integrations.
This avoids the need for calculation of higher-dimensional integrals.
The derivatives $dI/dR$ and $dj/d\Psi$ in equations~(\ref{abel})
and~(\ref{locVP}), respectively, are calculated using finite
differences. The derivative $d \Phi_{\rm tot}/dr$ in
equation~(\ref{jeanssol}) can be manipulated analytically so that
numerical differentiation is not needed. The accuracy of the code was
verified by reproducing analytically known results for special
potential-density pairs (e.g., Plummer models).


\newcommand{\figcapsb}{
(a; top panel) $V$-band surface brightness profile of {\omegaCen} in
mag arcsec$^{-2}$ versus radius in arcsec. Data points are from
various sources as described in the text. The curve is our best model
fit of equation~(\ref{nuker}), which has a central cusp slope $\gamma
= 0.05$. (b; bottom panel) Residuals of the fit in the top panel. The
small solid dots (black in the on-line version) are the HST star count
data from Paper~I. The bigger open symbols are from the compilation of
ground-based data in Trager \etal (1995), as described in the text:
GBANN, hexagons (green); DANN, circles (red); DSCAN, squares
(magenta); ADH-1792, triangles (cyan); ADH-1846, crosses
(blue).\label{f:sb}}

\begin{figure*}[t]
\epsfxsize=0.8\hsize
\centerline{\epsfbox{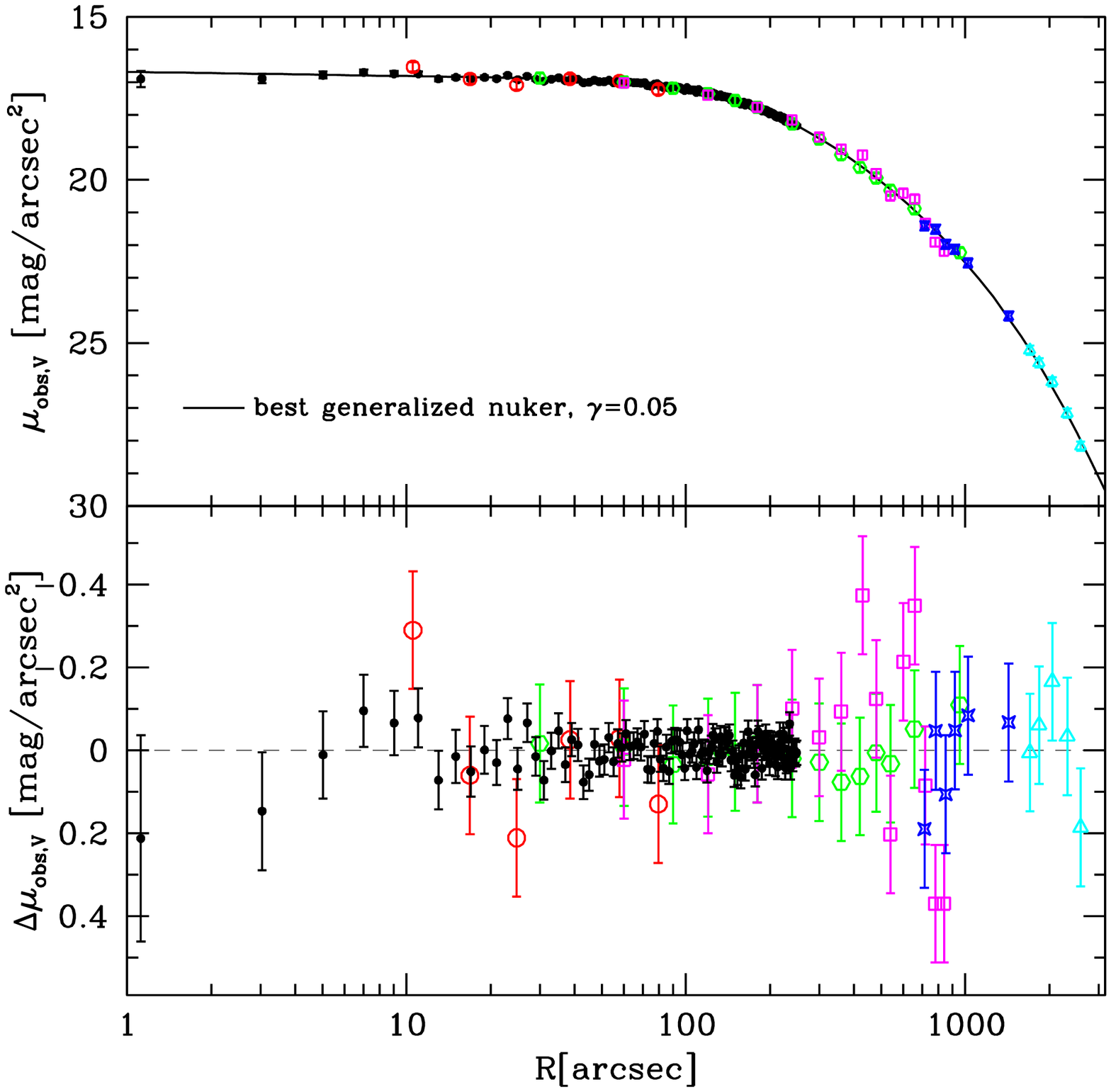}}
\figcaption{\figcapsb}
\end{figure*}


\section{Projected Density Profile Data}
\label{s:sbdata}

The primary input for our models is the projected radial density
profile. To characterize this profile over the full range of radii
spanned by {\omegaCen}, we have combined the ground-based measurements
at intermediate and large radii collected in Trager, King
\& Djorgovski (1995), with the new HST measurements at small radii
presented in Paper~I. Much of the measurements come in the form of
projected star count profiles $\nu_{\rm proj}(R)$. Our equations of
Section~\ref{s:modeling} are mostly expressed in terms of the observed
(or implied) surface brightness profile $\mu_{\rm obs}(R)$. The
profiles $\nu_{\rm proj}(R)$ and $\mu_{\rm obs}(R)$ are uniquely
related to each other (see eq.~[\ref{numu}] below), given our
assumption to neglect mass segregation among the luminous stars.

\subsection{Ground-based Data}
\label{ss:sbground}

Trager \etal compiled literature data from various sources. We
rejected from their compilation the data points to which they did not
assign the maximum weight (unity in their notation). This leaves only
the most reliable literature sources for \omegaCen, namely the
integrated-light photo-electric measurements of Gascoigne \& Burr
(1956) and Da Costa (1979), and the number density measurements of
King \etal (1968). All these authors presented averages for concentric
circular annuli. Da Costa also presented separate measurements using a
drift scan technique, which were subsequently calibrated to estimate
the profile along concentric circular annuli. King \etal presented
results for two separate plates from Boyden Observatory. We follow
Trager \etal by referring to the different data sets with the
abbreviations GBANN, DANN, DSCAN, ADH-1792, and ADH-1846; the latter
two abbreviations refer to the IDs of the photographic plates analyzed
by King et al. For our analysis we used the values provided by Trager
et al., who calibrated the published measurements to a $V$-band
surface brightness scale with a common zeropoint. We assigned to each
data point an error bar of $0.142$ mag, based on the procedure and
analysis of McLaughlin \& van der Marel (2005; see their
Table~6). This corresponds to the scatter in the data with respect to
a smooth curve.

\subsection{HST Data}
\label{ss:sbHST}

Paper~I presented number density measurements for concentric circular
annuli based on HST data. Considerable effort was spent to accurately
determine the cluster center, and to correct the results for the
effects of incompleteness. Error bars were determined from the Poisson
statistics of the counted stars. Paper~I discussed number density
profiles for various magnitude bins, and showed that there is no
strong dependence on magnitude. For the analysis presented here we
have used the profile for all stars in the HST images with $B$-band
instrumental magnitude $< -10$ (this is defined in Paper~I and
corresponds to stars measured with signal-to-noise ratio $S/N \gta
100$). The choice of magnitude range was motivated by the desire to
maximize the number of stars (i.e., to minimize the error bars on the
density profile) while maintaining a similar mean magnitude ($\sim
-12$) as for the sample of stars for which high-quality proper motions
are available from Paper~I.

To extend the profile from the Trager \etal compilation to the smaller
radii accessible with the HST data it is necessary to calibrate the
projected number density $\nu_{\rm proj}(R)$ to a projected surface
brightness $\mu_{{\rm obs,}V}(R)$. It follows from
equations~(\ref{obsmag}), (\ref{radscale}), (\ref{nuj}), and
(\ref{flam}) that
\begin{equation}
\label{numu}
  \mu_{{\rm obs,}V} [{\rm mag} \> {\rm arcsec}^{-2}] = 
     -2.5 \log (\nu_{\rm proj} [{\rm arcsec}^{-2}]) + Z ,
\end{equation}
where the zeropoint $Z$ is equal to
\begin{equation}
  Z = 21.572 + M_{\odot, V} + 
      2.5 \log \lbrace (f/\lambda[L_{\odot}]) 
                       (\pi D[{\rm kpc}] / 648 )^{2} \rbrace ,
\end{equation}
where $M_{\odot, V} = 4.83$ (Binney \& Merrifield 1998). Although
$(f/\lambda[L_{\odot}])$ can in principle be estimated, we have found
it more reliable to calibrate $Z$ so that the surface brightness
profiles from Trager \etal and Paper~I agree in their region of
overlap. We have done this by including $Z$ as a free parameter in the
fits described below, which yields $Z = 18.21$. For a canonical
distance $D = 4.8 \kpc$ (vdV06) this implies that
$(f/\lambda[L_{\odot}]) = 1.0$. With this $Z$ there is excellent
agreement between the different datasets, as shown in
Figure~\ref{f:sb}a.

NGB08 studied the surface brightness profile of unresolved light in
{\omegaCen} using HST images. In principle, the advantage of using
unresolved light over star counts is that one can trace the cumulative
effect of all the unresolved faint stars, rather than counting only
the less numerous resolved ones. However, we showed in Paper~I that
even at HST resolution, the unresolved light is dominated by scattered
light from the PSF wings of only the brightest stars. So in reality,
the inherent statistics of unresolved light measurements are poorer
than those of star count measurements. Moreover, we showed in Paper~I
that the combined photometric and kinematic center of {\omegaCen} is
actually $12''$ away from the position where NGB08 identified it to
be. There is no simple way to transform their measurements over
concentric annuli to a different center position. For these reasons,
we have not included the NGB08 surface brightness datapoints in the
dataset used for the dynamical modeling.


\newcommand{\figcapsbcen}{
$V$-band surface brightness profile of {\omegaCen} in mag
arcsec$^{-2}$ versus radius in arcsec. The data and the symbols are
the same as in Figure~\ref{f:sb}a, but this figure focuses more on the
central region. The solid curve (black in the on-line version) is the
overall best model fit of equation~(\ref{nuker}), which has a central
cusp slope $\gamma = 0.05$. The long-dashed curve (blue) is the best
fit when the cusp slope is fixed a priori to $\gamma = 0$. The dashed
(cyan) and dotted (magenta) curves are the Wilson and King models,
respectively, presented in McLaughlin \& van der Marel (2005). These
were tailored to fit the compilation of ground-based data in Trager
\etal (1995).\label{f:sbcen}}

\begin{figure*}[t]
\epsfxsize=0.8\hsize
\centerline{\epsfbox{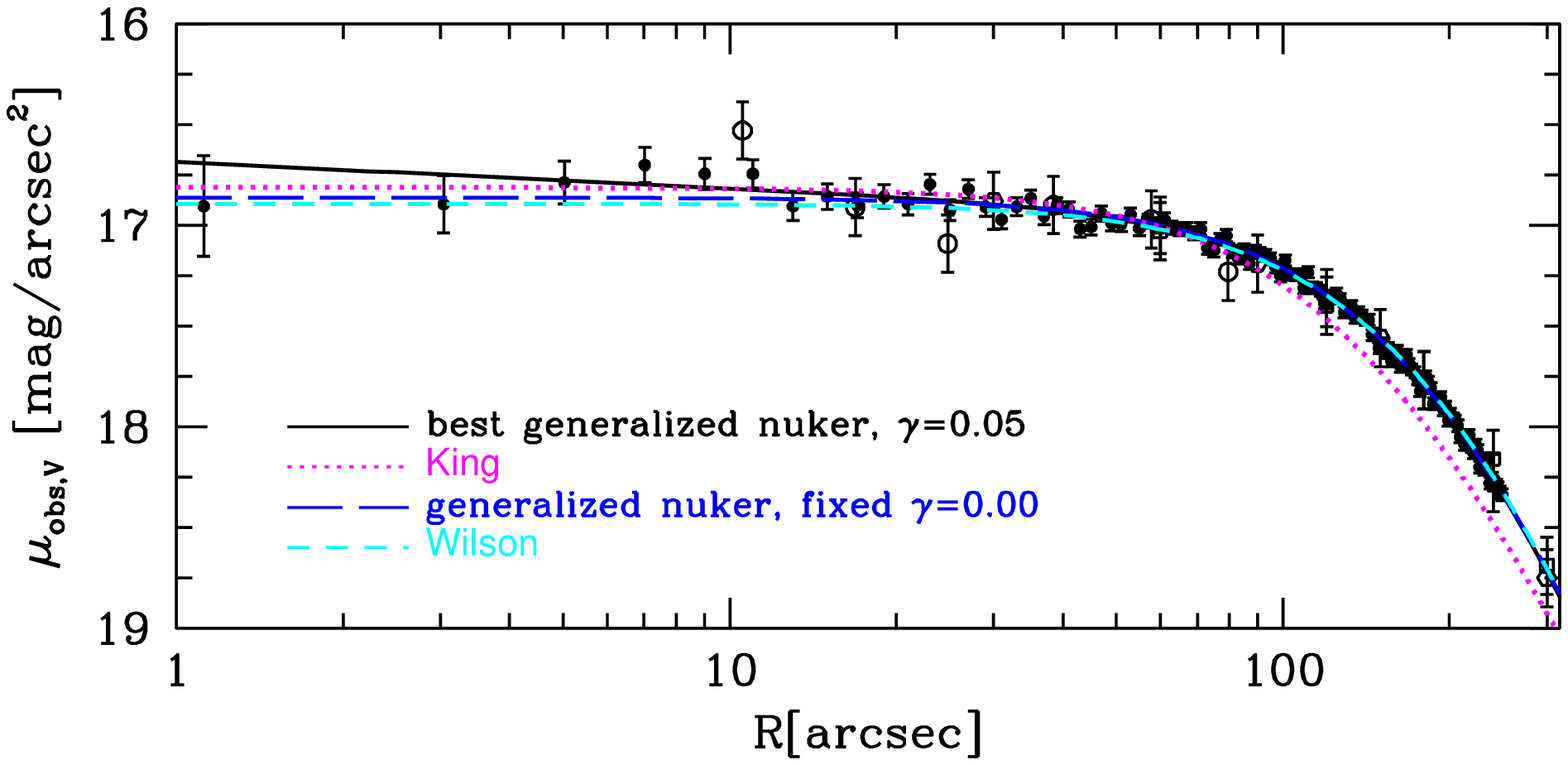}}
\figcaption{\figcapsbcen}
\end{figure*}


\section{Projected Density Profile Models}
\label{s:sbmodel}

It is important for the modeling to have a smooth representation of
the projected intensity profile $I(R)$, since the derivative $dI/dR$
is needed in equation~(\ref{abel}) for the calculation of $j(r)$. For
the dynamical modeling we have used a parametric approach, although
this can in principle be done non-parametrically (e.g., Gebhardt \&
Fischer 1995). We have fit the data using a smooth function of the
form
\begin{eqnarray}
\label{nuker}
 I_{\rm obs} (R) &=& 
           I_b \> 2^{(\eta-\gamma)/\alpha} \> (R/b)^{-\gamma} \nonumber\\
      & & \left [ 1+(R/b)^{\alpha} \right ]^{-(\eta-\gamma)/\alpha}
          \left [ 1+(R/c)^{\delta} \right ]^{-(\epsilon-\eta)/\delta} .
\end{eqnarray}
This functional form has no particular physical significance, but it
is useful because it can fit a wide range of observed profiles. The
profile has a power-law cusp $I \propto R^{-\gamma}$ at radii $R \ll
b$. It then breaks to a logarithmic slope $\eta$ at $R \approx b$, and
further to a logarithmic slope $\epsilon$ at $R \approx c > b$. The
parameters $\alpha$ and $\delta$ measure the sharpnesses of the
breaks. The intensity $I_b$, corresponding to a magnitude $\mu_b$
through equation~(\ref{obsmag}), sets the overall intensity scale.
The profile is similar to a so-called ``nuker'' profile (Lauer \etal
1995), but with two breaks instead of just one. We will refer to it as
a generalized nuker profile.

The best fit to the dataset presented in Section~\ref{s:sbdata} was
determined using a Levenberg-Marquardt iteration scheme (Press \etal
1992). It has parameters $\mu_b = 17.74$, $b = 183.1''$, $c = 2158''$,
$\gamma = 0.05$, $\eta = 2.37$, $\epsilon = 10.00$ (the maximum value
allowed during the iteration), $\alpha = 2.29$, and $\delta =
1.70$. The fit and its residuals are shown in Figure~\ref{f:sb}. The
quality of the fit is excellent. For the subset of the data from
Paper~I, the $\chi^2 = 130.5$ for 125 datapoints with an RMS residual
of 0.04 mag. The total $\chi^2_{\rm min} = 169.3$ for all 159
datapoints, corresponding to 151 degrees of freedom. This is
acceptable at the $\sim 1 \sigma$ level for a $\chi^2$
distribution. The residuals appear generally consistent with random
scatter in line with the error bars, with little evidence for
systematic trends.

The best fit has a central cusp slope of $\gamma = 0.05$. To determine
the statistical error on this value we also performed a range of fits
in which $\gamma$ was fixed a priori, while all other parameters were
varied to optimize the fit. The central structure of our best fit for
$\gamma=0$ is shown in Figure~\ref{f:sbcen} for comparison to our
overall best fit.  The curve of $\Delta \chi^2
\equiv \chi^2 - \chi^2_{\rm min}$ as function of $\gamma$ was used to
derive the error bar on $\gamma$, which yields $\gamma = 0.05 \pm
0.02$.

When using a generalized nuker fit, a constant surface brightness core
is inconsistent with the data at $\sim 2.5\sigma$ confidence. However,
this parameterized approach may not yield an unbiased estimate of the
asymptotic slope $\gamma$ at the smallest radii. The data at all radii
are given equal weight in the $\chi^2$ of the fit, but it is obviously
only the data near the very center that contain most of the relevant
information on $\gamma$. Also, the two breaks at radii $b$ and $c$ in
the generalized nuker profile fit correspond to the breaks associated
with the traditional core and tidal radii of {\omegaCen}. The
parameterization does not have sufficient flexibility to reproduce any
possible change in profile slope well within the core radius ($r_0 =
141_{-13}^{+7}$ arcsec; McLaughlin \& van der Marel 2005).

An alternative and simpler approach to determine $\gamma$ is to fit a
straight line to the $(\log R,\log I)$ data at $R \leq R_{\rm
max}$. For $R_{\rm max} = 15''$, this yields $\gamma = 0.00 \pm
0.07$. The values of $\gamma$ inferred for other values of $R_{\rm
max}$ in the range $0$--$20''$ are in statistical agreement with
this. This analysis indicates that the central data points are
consistent with a flat core.

An additional assessment of models with a flat core can be obtained
from comparison of the new data with previously published
models. McLaughlin \& van der Marel (2005) presented detailed surface
brightness fits for a large sample of galactic and extragalactic
globular clusters. They considered three types of models with
decreasing amounts of diffuse outer structure, the power-law model,
the Wilson (1975) model, and the King (1966) model. The power-law
models are similar to the parameterizations that we have used here,
but with a more restricted subset of the parameters (namely,
$\gamma=0$, $\alpha=2$, and $c=\infty$ were kept fixed). The models
from Wilson and King are both based on a parameterized isotropic
phase-space distribution function. The Wilson prescription differs
from the better known King models in that it uses a slightly different
lowering of the exponential energy distribution, so as to produce a
model that has a more extended but still finite outer
structure. McLaughlin \& van der Marel discussed {\omegaCen} as a
special example (their section~4.2.1) and found that it is well fit by
a Wilson model, but less so by a King model. The central structure of
both models is shown in Figure~\ref{f:sbcen}. The Wilson model is very
similar to our best fit model with fixed $\gamma=0$. The King model is
also similar at radii $R \lta 100''$, but it under-predicts the
brightness at intermediate radii (before improving again at larger
radii, not shown in Figure~\ref{f:sbcen}).

The Wilson and King models shown in Figure~\ref{f:sbcen} are the ones
that were previously fit by McLaughlin \& van der Marel to the Trager
\etal data. At small radii these models represent an extrapolation of
the ground-based data under the assumption of a constant surface
brightness core. Although the models were not fit to the HST data,
they do provide a perfect fit to the two innermost HST star count data
points. By contrast, our own best generalized nuker model fit with
$\gamma = 0.05$ over-predicts both of the innermost datapoints by
about $1\sigma$. The fact that $\gamma = 0.05$ is preferred in our
fits over $\gamma = 0$ is therefore because it fits better the data at
radii $R \approx 10''$, and not because it provides the best fit
closest to the center. 

NGB08 inferred from their HST study of integrated light in {\omegaCen}
that it has a central cusp slope $\gamma = 0.08 \pm 0.03$. At first
glance, this result appears consistent with that from our generalized
nuker fit. However, since they did not measure the brightness profile
around the same cluster center, the agreement between our inferred
cusp slopes is likely coincidental.

In summary, we conclude that the central cusp slope of {\omegaCen} is
$\gamma \lta 0.07$ at $\sim 1\sigma$ confidence. That is, we interpret
the $\gamma = 0.05 \pm 0.02$ from our generalized nuker fit as an
upper limit. We do this because it over-predicts the central two data
points, and because other statistics do not show evidence for a
non-zero slope. In the dynamical analysis that follows we specifically
consider two generalized nuker fits that span the range of
possibilities: models with a shallow cusp ($\gamma = 0.05$) and models
with a constant density core ($\gamma = 0$). We will refer to these as
cusp models and core models, respectively. In all calculations we have
transformed the observed intensity $I_{{\rm obs},V}(R)$ to an
extinction-corrected intensity $I_V(R)$ using equation~(\ref{projint})
with $A_V = 0.372$ for {\omegaCen} (McLaughlin \& van der Marel 2005).

\section{Kinematical Data}
\label{s:kinematics}

To constrain the parameters of the dynamical models we need to use
both line-of-sight and proper motion data at a range of radii. For
this we have combined ground-based line-of-sight and proper motion
data at intermediate and large radii with the new HST proper motion
data at small radii presented in Paper~I.

We construct spherical models to interpret the data. These models
yield only a single radial profile for each projected quantity, with
no azimuthal variations from the observed major axis of {\omegaCen} to
its observed minor axis. To get accurate results for the model
parameters, it is important that these models be used to reproduce the
{\it average} radial variations seen in the data. This is achieved by
using for each projected quantity that is fit the average of the
measurements along concentric annuli. This is indeed what we extracted
for the projected density in Section~\ref{s:sbdata}, and we therefore
do the same here for the projected velocity distributions. When
averages are taken over annuli, all the odd moments of the velocity
distribution are zero. Of primary interest here are the second
moments, which yield the RMS velocity $\slos$ and the RMS proper
motions $\Spmr$ and $\Spmt$.

In reality, variations do exist between observed quantities along the
major and minor axis. vdV06 discussed this in detail for the existing
ground-based data. In Appendix~B we address this for the new HST
proper motion data. In particular, the RMS proper motions are not the
same along the major and minor axes, and the RMS proper motions in the
major and minor axis directions on the sky are not the same. Also, the
mean proper motion in the transverse (pmt) direction is not zero
(vdV06; Section~\ref{sss:HSTrot}; Appendix~C). By modeling these
issues it is possible to constrain, e.g., the cluster inclination and
its rotation rate. However, it is not essential to model these issues
to achieve a good understanding of the central mass distribution,
which is the main goal here.


\newcommand{\figcapkin}{
Observed kinematical quantities for {\omegaCen} as function of
projected radius $R$ from the cluster center. Data points are based on
analysis of various sources described in the text: circles
(magenta in the on-line version) are based on vdV06, which is itself a
compilation of line-of-sight data from Suntzeff \& Kraft (1996), Mayor
\etal (1997), Reijns \etal (2005), and Gebhardt \etal (in prep.), and
proper motion data from van Leeuwen \etal (2000); triangles (cyan) are
based on NGB08; stars (blue) are based on Sollima \etal (2009);
squares (green) are based on Seitzer (1983); crosses (red) are based
on Scarpa \etal (2003); and small dots (black) are based on the HST
data from Paper~I. Some error bars are of similar size as the plot
symbols. Data from all ground-based sources except NGB08 were
multiplied by a factor 1.023 to correct to same the characteristic
main-sequence stellar mass as for the HST sample (see
Section~\ref{sss:massseg}). (a; top panel) RMS line-of-sight velocity
$\slos$ in km/s (data indicated with open symbols). (b; middle panel)
RMS proper motion $\Spm$ in mas/yr, averaged over all directions in
the plane of the sky, as defined by equation~[\ref{pmtot}] (data
indicated with closed symbols). (c; bottom panel) Ratio $\Spmt/\Spmr$
of the RMS motions in the tangential and radial proper motion
directions (data indicated with closed symbols). The RMS quantities in
each dataset are averages over adjacent circular annuli. The extent of
the annulus for data point $i$ can be assessed by taking half the
distance between adjacent datapoints, $\Delta R_i \approx (R_{i+1} -
R_{i-1})/2$.\label{f:kin}}

\begin{figure*}[t]
\epsfxsize=0.8\hsize
\centerline{\epsfbox{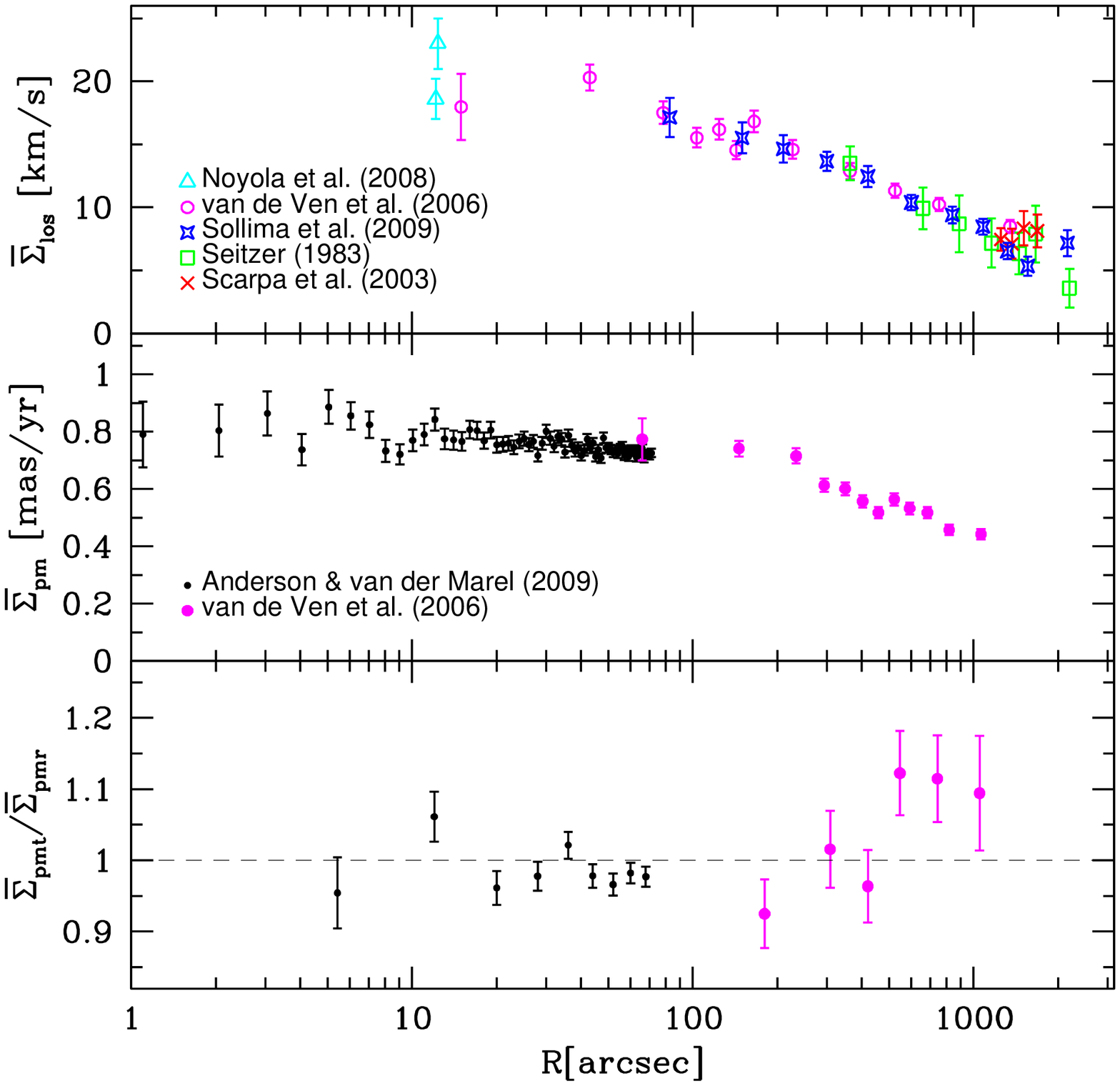}}
\figcaption{\figcapkin}
\end{figure*}


\subsection{Ground-based Data}
\label{ss:ground}

\subsubsection{Line-of-Sight Kinematics}
\label{sss:groundlos}

vdV06 compiled line-of-sight velocity data of individual stars in
{\omegaCen}. The original data were obtained by Suntzeff \& Kraft
(1996), Mayor \etal (1997), Reijns \etal (2005), and Gebhardt \etal
(in preparation). vdV06 performed various operations and cuts to
homogenize and correct the data as necessary for dynamical
modeling. This included: improved removal of cluster non-members;
removal of stars with low enough brightness or large enough velocity
error bars to make their reliability suspect; application of additive
velocity shifts to correct for offsets in velocity zero-points between
data sets; multiplicative rescaling of the velocity error bars where
necessary, as dictated by actual measurements of scatter between
repeat measurements; subtraction of the systemic line-of-light
velocity; and subtraction of the apparent solid-body rotation
introduced into the velocity field through the combination of systemic
transverse motion and large angular extent (``perspective
rotation''). This produced a final homogenized sample of line-of-sight
velocities for 2163 stars with uncertainties below $2.0 \kms$.  

Glenn van de Ven kindly made the homogenized sample from vdV06
available to us for use in our data-model comparisons. To determine
the profile of $\slos(R)$ we binned the stars in radius and then used
the statistical methodology described in Appendix~A. Radii were
calculated with respect to the center determined in Paper~I, which
differs by $15.5''$ from the center used by vdV06 (which was
determined by van Leeuwen \etal 2000). We chose a central bin of
$20''$ radius (with 25 stars) for maximum resolution near the center,
and then chose the remaining bins to contain 200 stars each.

Sollima \etal (2009) obtained line-of-sight velocities of 318 cluster
members at large radii, which they merged with observations closer to
the center by Pancino \etal (2007). This yielded a homogeneous sample
of 946 cluster members, from which they derived the profile
$\slos(R)$. Seitzer (1983) presented the $\slos(R)$ profile derived
from 118 clusters members with known velocities at a range of
radii. We included the published profiles from both studies in our
data-model comparisons. For each study we averaged the first two bins
in the profile together, since the first bin had a large error bar and
did not provide spatial resolution that competes with the vdV06
compilation.

Scarpa \etal (2003) obtained line-of-sight velocities of 75 cluster
members at large radii. From this they determined the velocity
dispersion $\sigma_{\rm los}$ in four radial bins. The stars were
limited to the West side of the cluster, so the results do not
represent averages over a complete annulus on the sky. Nonetheless, we
included these observations in our data-model comparisons. We
subtracted in quadrature from the published $\sigma_{\rm los}$ values
the average measurement error of $1 \kms$. To obtain $\slos$ we added
in quadrature $V^2/2$, where $V$ is the rotation velocity amplitude
quoted in Scarpa \etal (2003); the factor $1/2$ is the average of a
squared sinusoidal variation over a full range of angles.

The profiles obtained from the Sollima \etal (2009), Seitzer (1983),
and Scarpa \etal (2003) observations do not pertain to the same center
as that inferred in Paper~I. However, the smallest radial bin that we
use from these studies is the 4 arcmin diameter combined central bin
of Sollima et al. A center offset of $15.5''$ (as applicable to the
Sollima \etal and Scarpa \etal results, who used the van Leeuwen \etal
2000 center) should not significantly affect statistics calculated
over bins that extend to such large distances from the center.

At small radii we used as additional data points the integrated-light
measurements from NGB08. They inferred velocity
dispersions over square fields of $5'' \times 5''$. Because of the
significant size of these fields, we did not include the impact of PSF
convolution in our model predictions for these data. We placed the
measurements at the distance from the cluster center inferred in
Paper~I, which is $12''$ for both fields. We cannot use the data to
calculate a true average over an annulus. However, we do have two
measurements at the same distance from the center at a $60^{\circ}$
angle. Including both measurements in our model fits is in low-order
approximation similar to fitting the average over an annulus.

NGB08 did not measure the mean velocity over their fields as compared
to the systemic velocity. Hence, their measurements are dispersions
$\sigma_{\rm los}$ rather than RMS velocities $\slos$. However, it is
known from previous observations (e.g., vdV06) that $V_{\rm
los}/\sigma_{\rm los}$ in {\omegaCen} decreases towards the
center. This ratio is small enough at the position of the NGB08 fields
that $\slos \approx \sigma_{\rm los}$ to high accuracy (see
Section~\ref{sss:HSTrot} for a quantitative estimate).

Figure~\ref{f:kin}a shows the collected $\slos$ data from all the
sources, as function of projected distance $R$ from the cluster
center. The different data sets match well in their regions of
overlap.

\subsubsection{Proper Motion Kinematics}
\label{sss:groundpm}

Ground-based proper motion data are available from van Leeuwen \etal
(2000). vdV06 performed various operations and cuts to optimize and
correct the data as necessary for dynamical modeling. This included:
improved removal of cluster non-members; removal of stars with low
enough brightness, large enough velocity error bars, or neighbors that
are nearby enough to make their reliability suspect; and subtraction
of the apparent solid-body rotation introduced by perspective
rotation, and other spurious solid body rotation components. This
produced a final sample of proper motions for 2295 stars with
uncertainties below $0.2 \masyr$ (i.e., $4.6 \kms$ at the canonical
distance $D = 4.8 \kpc$).

Glenn van de Ven kindly made the homogenized sample from vdV06
available to us for use in our data-model comparisons. Proper motions
were given along the major and minor axis directions. For each star we
transformed this into proper motions in the radial (pmr) and
tangential (pmt) directions. We then determined the profiles of
$\Spmr(R)$ and $\Spmt(R)$ by binning the stars in radius and using the
statistical methodology described in Appendix~A. Radii and proper
motion coordinate transformations were calculated with respect to the
center determined in Paper~I. We chose a central bin of $90''$ radius
(with 30 stars) for maximum resolution near the center, and then chose
the remaining bins to contain 200 stars each. We followed vdV06 by
excluding the 65 outermost stars with radii $R \sim 20$--30 arcmin.

For fitting models to the data we always use the quantities $\Spmr$
and $\Spmt$. However, for visualizing the results we have found it
more convenient to use the quantities ${\Spm}$ and $\Spmt/\Spmr$. The
RMS proper motion ${\Spm}$ averaged over all directions in the plane
of the sky can be calculated directly through its definition in
equations~(\ref{pmtot}) and~(\ref{pmunit}). In practice, we found it
more convenient to calculate it by applying the statistical
methodology described in Appendix~A on the {\it combined} array of the
pmr and pmt data points. For visualization of the ratio $\Spmt/\Spmr$
we adopted a radial binning scheme with 400 stars in each bin, to
obtain smaller and more useful error bars.

Figure~\ref{f:kin}b shows the $\Spm$ data as function of projected
distance $R$ from the cluster center, and Figure~\ref{f:kin}c shows
the $\Spmt/\Spmr$ data. The velocity distribution is close to
isotropic near the center, but with some radial anisotropy. There is a
significant increase in tangential anisotropy at a few core radii. The
latter is mostly due to the contribution of rotation to the second
moment $\Spmt$. The anisotropy in the proper motion {\it dispersions}
is much less (King \& Anderson 2002).

\subsubsection{Dependence of Kinematics on Stellar Mass}
\label{sss:massseg}

Luminous stars in a cluster undergo two distinct effects as a result
of two-body relaxation: mass segregation and energy equipartition.
These effects are intimately connected through basic dynamical theory,
but from an observational perspective it is useful to think of them as
separate effects. Mass segregation causes stars of different masses to
have slightly different spatial distributions. In
Section~\ref{s:modeling} we assumed that this could be neglected when
parameterizing the run of mass-to-light ratio with radius, and when
connecting star count data at small radii to surface brightness data
at large radii. This assumption was motivated by the absence of
high-quality data to constrain the amount of mass segregation over the
full scale of the cluster, and by the desire to minimize the number of
a priori model assumptions. Energy equipartition causes stars of low
mass to move faster than those of high mass. We measured in Paper~I
that low-mass stars in {\omegaCen} do indeed move faster than
high-mass stars, albeit by less than would be predicted for complete
energy equipartition ($v \propto m^{-0.5}$). Since we have a direct
measurement of this, there is no reason to ignore this effect in the
modeling. We therefore correct for it when comparing the observed
kinematics from different data sets.

The available discrete ground-based line-of-sight and proper motion
data all pertain to relatively bright stars. The apparent magnitudes
of these stars place them on the giant or sub-giant branch. Since
stellar evolution proceeds relatively rapidly after the main-sequence
turn-off, these stars all have essentially the same mass as the
main-sequence turn-off.  By contrast, the stars in our new HST proper
motion sample from Paper~I (discussed in detail in
Section~\ref{ss:HSTkin} below) extend from about the main-sequence
turn-off to $\sim 3$ magnitudes below that. From the analysis of
Paper~I it follows that the RMS proper motion of stars at the average
magnitude of the HST sample is $\sim 2.3$\% higher than that at the
main-sequence turn-off.

For a fair comparison to the new HST proper motion data, we took the
kinematics from the discrete velocity studies in
Sections~\ref{sss:groundlos} and~\ref{sss:groundpm} and multiplied
them by a factor $1.023$. This implies that we use the observed
(sub-)giant star kinematics to estimate the kinematics of lower-mass
stars in the same general population. We assume implicitly that the
variation of RMS kinematics with mass is the same everywhere in the
cluster (although it was only measured near the cluster center), and
that it applies also in the line-of-sight direction (although it was
only measured in the proper motion direction). In reality, one would
expect the relation between mass and velocity dispersion to change
with radius, since relaxation proceeds faster near the cluster center
than further out. This was neglected in the present context by
applying the same small correction throughout the cluster. This was
motivated in part by the fact that our study focuses mostly on the
central part of the cluster anyway, but also by the fact that we
didn't actually detect a difference in equipartition in Paper~I
between the different (central and major axis) fields for which we
obtained proper motions. We did not apply the multiplicative
correction to the NGB08 data. Their measurements are based on
integrated light, excluding the brightest regions in their fields. It
is therefore not obvious that their measurements would be dominated by
(sub-)giant stars. Either way, a $2.3$\% correction would be much less
than the error bars on their datapoints.

In summary, we are using our Jeans models to fit the kinematics of
stars with masses typical of our HST proper motion sample. We extend
our actual HST kinematics to larger radii using ground-based
kinematics that have been corrected to the same characteristic
mass. We model the cluster density profile using HST star counts for a
sample that is also centered on the same characteristic mass (see
Section~\ref{ss:sbHST}). We make the simplifying assumption that there
is no mass segregation among the luminous stars only so that we can
extend the HST density profile to larger radii, where it can be tied
to ground-based star-count data and integrated photometry that are
based largely on giant stars (see Section~\ref{ss:sbground}). Since
this approximation does not affect the very central region of the
cluster, it should have little effect on our results for the central
mass distribution. In fact, the multiplicative correction of $1.023$
discussed above also has little effect on this. We verified explicitly
that none of the main conclusions of our paper change at a level that
exceeds the formal uncertainties when the multiplicative correction is
omitted.

\subsection{HST Proper Motion Data}
\label{ss:HSTkin}

\subsubsection{Sample selection}
\label{sss:HSTsample}

In Paper I we derived proper motions for stars observed in two fields
observed by HST, a ``central field'' on the cluster center, and a
``major axis field'' positioned adjacent to it in non-overlapping
fashion roughly along the cluster major axis. Here we use the
``high-quality'' sample presented in Paper~I, which contains the
non-saturated stars that are isolated enough and bright enough
($B$-band instrumental magnitude $< -11$) for a particularly reliable
proper motion measurement. This sample has 53382 stars in the central
field and 19593 stars in the major axis field. In this sample, 95.1\%
of the stars have proper motion errors below $0.2 \masyr$. Our cut in
proper motion accuracy is therefore not very different from that
applied by vdV06. However, the median proper motion error per
coordinate in the final sample of vdV06 is $\sim 0.14 \masyr$, whereas
it is a factor two smaller, $0.07 \masyr$, for our sample. We note
that our use of only the high-quality sample is rather
conservative. Even the remainder of the sample of Paper~I is quite
accurate, and could well have been used to constrain the cluster
dynamics.

The central HST field covers radii $R \leq 147.4''$, while the major
axis field covers radii $99.7'' \leq R \leq 347.1''$. The central
field has complete position angle coverage out to $R = 71.7''$ (the
radius of the largest enclosed circle). For $71.7'' \leq R \leq
147.4''$ the coverage over the range of position angles becomes
increasingly sparse. The major axis field is restricted at all radii
to position angles within $48.5^{\circ}$ of the major axis. For the
analysis of the present paper it is important to have coverage of all
position angles at a given radius, so that average kinematical
quantities over circular annuli can be calculated. For this reason we
have used only the 25194 stars at $R \leq 71.7''$ to constrain the
dynamical models.

In our analysis we ignore the 47781 stars with high-quality HST proper
motions at radii $71.7'' \leq R \leq 347.1''$. We do discuss relevant
quantities derived from these data in Appendix~C. More sophisticated
axisymmetric modeling techniques (such as those in vdV06; van den
Bosch \etal 2006; and Chaname, Kleyna, \& van der Marel 2008) that
might be applied in the future probably would not want to exclude
these data in their analysis. By contrast, to include these data in
our spherical models would require estimates of the ratios ${\cal R}$
of the average of either $\Spmr$ or $\Spmt$ over a restricted range of
position angles, to its average over a circle. We have experimented
with such estimates, but concluded that the approximations that need
to be made introduce sufficient uncertainties that these data do not
really help to constrain our models. Conversely, we have found no
evidence that inclusion of the HST data at $71.7'' \leq R \leq
347.1''$, combined with reasonable estimates for the ratios ${\cal
R}$, would alter in any way the conclusions draw below (see
Appendix~C).

To reject cluster non-members we created scatter plots of total
two-dimensional proper motion $|{\vec p}|$ versus radius $R$, where
${\vec p}$ is the proper motion vector of an individual star.  Such a
plot reveals a sparse population of field stars at large values of
$|{\vec p}|$, in addition to the numerous cluster stars at low $|{\vec
p}|$. We rejected those stars from the sample that reside at $R >
10''$ and for which $|{\vec p}| > 2.8 - (r/500'') \masyr$. At the
radii for which we have HST proper motions, this equation is a good
approximation to the escape velocity curve for a spherical model with
a canonical distance and mass-to-light ratio ($D=4.8 \kpc$,
$\Upsilon_{\rm V} = 2.5$; vdV06).  With this criterion, only 27
non-member stars were identified at $R \leq 71.7''$, the closest of
which resides at $R=21.4''$. Since the presence of an IMBH may result
in the presence of rapidly moving stars at small radii, we did not
reject any stars at $R \leq 10''$. The wings of the observed velocity
distribution at small radii are discussed in detail in
Section~\ref{ss:isoGH} below.

Our final HST proper motion sample for the dynamical modeling consists
of 25167 stars. This is a factor $\sim 11$ larger than the number of
stars with proper motions in the vdV06 sample. Most of their stars
reside at larger radii than those sampled here. At the radii $R
\leq 71.7''$ of our sample, vdV06 had only some two dozen stars
with proper motions, and $\sim 270$ stars with line-of-sight
velocities. The HST data therefore provide a major advance for
constraining dynamical models, especially as it pertains to the
gravitational influence of a possible IMBH at small radii.

\subsubsection{RMS Proper Motions}
\label{sss:HSTRMS}

For each star in the HST sample we decomposed the observed proper
motion vector ${\vec p}$ into components in the pmr and pmt
directions, respectively. We then binned the stars in radius and
derived the profiles $\Spmr(R)$, $\Spmt(R)$, and $\Spm(R)$. We did
this in similar fashion as for the ground-based proper motion data
discussed in Section~\ref{sss:groundpm}, using the statistical
methodology described in Appendix~A.

Figure~\ref{f:kin}b shows the $\Spm$ data as function of projected
distance $R$ from the cluster center. The results match well with
those from vdV06 in the region where the data sets overlap. For the
individual data points of $\Spmt$, $\Spmr$ and $\Spm$ we used binning
in annuli that are $1''$ wide. For the central aperture we used a
circle of radius $1.55''$, which gives an average radius $R=1.1''$ for
the datapoints enclosed by it. This central aperture is larger than
the $\sim 1''$ uncertainty in the position of the cluster center
derived in Paper~I. This uncertainty should therefore not
significantly affect the inferred kinematics. The adopted binning
gives acceptably small error bars in all apertures, without losing too
much spatial resolution near the center.

One immediate result from the inferred $\Spm(R)$ profile is that there
is no strong increase towards the center, consistent with arguments
presented in Paper~I. A straight line fit to the $(R,\Spm)$ data for
$R \leq R_{\rm max}$, with $R_{\rm max} = 15''$, yields a slope of
$-0.06 \pm 0.08 \kms \> {\rm arcsec}^{-1}$.  The slopes inferred for
other values of $R_{\rm max}$ in the range $0$--$20''$ are
statistically consistent with this. This analysis indicates that the
central data points are consistent with a flat profile of RMS velocity
as function of radius.

The error bars on the observed RMS velocities can be reduced by
increasing the amount of binning. For the ratio of $\Spmt/\Spmr$ shown
in Figure~\ref{f:kin}c we used binning in annuli of $8''$ wide, so
that even small deviations from isotropy can be measured. As for
$\Spm$, the results match well with those from vdV06 in the region
where the data sets overlap. The average over all HST data points is
$\Spmt/\Spmr = 0.983 \pm 0.006$. Hence, there is a very small, but
significant, amount of radial anisotropy in the central arcminute of
{\omegaCen}.

\subsubsection{Proper Motion Rotation}
\label{sss:HSTrot}

Our data calibration of Paper~I (see discussion in Section 3.6.4 of
that paper) removed by necessity any possible solid-body rotation
component from the data. This is because we did not have access to
stationary background sources in the HST fields, as would be required
to measure absolute motions. This calibration has the advantage that
it automatically removes any perspective rotation present in the actual
proper motions (however, this is negligible for the small radii $R
\leq 71.7''$ sampled by our data anyway; $\leq 0.0035 \masyr = 0.08
\kms$ at the canonical $D = 4.8 \kpc$). But of course, it has the
disadvantage that it also removes a solid-body rotation fit to the
actual rotation field of the cluster. This is a fundamental limitation
of the data. However, we discuss below that this has only negligible
importance for our models.

In Paper~I, we also used a ``local correction'' that removed all other
mean systematic motions (as opposed to random motions) from the proper
motion data. However, that was by choice, and not by necessity. The
motivation for this is to calibrate out small time-variations in the
higher-order geometric distortion terms. This improves the
measurement of the random motions in the cluster. However, one would
of course not want this to remove an important signal that is present
in the data, and in particular, any possible differential rotation in
the plane of the sky (i.e., the part of the rotation field that
remains after subtraction of a solid-body fit). We tested for such
differential rotation in Paper~I, and found that it is negligible in
the central HST field. Proper motion rotation curves (the mean
tangential motion as function of radius) derived from data with and
without the local correction agree to within $1 \kms$ at all
radii. Therefore, solid-body rotation is the {\it only} unrepresented
rotation component that could affect the modeling.

In principle, the loss of information on solid-body rotation can be
partly recovered as in vdV06. They use the fact that for an
equilibrium axisymmetric system the equation
\begin{equation}
  V_{\rm los} = 4.7404 \> D[\kpc] \> \tan i \> {\cal V}_{\rm y}
\end{equation}
must hold everywhere on the projected plane of the sky. Here $i$ is
the inclination, $y$ the projected minor axis direction, and the
symbol ${\cal V}$ is used to distinguish a mean proper motion from a
mean velocity $V$. However, to use this equation with accuracy for our
HST proper motions would require many more stars with line-of-sight
velocities than are actually available at $R \leq 71.7''$.

The pmr direction is perpendicular to the direction of absolute
rotation, and our estimates of $\Spmr$ are therefore not affected by
our inability to measure absolute rotation. Also, in an equilibrium
system there cannot be mean streaming in the pmr direction, when
averaged along a circle. However, our calculated values of $\Spmt =
({\cal V}_{\rm pmt}^2 + \Sigma_{\rm pmt}^2)^{1/2}$ will be
systematically low by a multiplicative factor $g \leq 1$, because we
are unable to include any actual solid-body contribution from ${\cal
V}_{\rm pmt}^2$. The proper motion data presented by vdV06 {\it did}
include measurements of absolute rotations in the plane of the
sky. Figure~\ref{f:vsig} shows for their data the ratios ${\cal
V}_{\rm pmt}/\Sigma_{\rm pmt}$ and $g \equiv \Sigma_{\rm pmt} /
\Spmt$, where the quantities in the nominator and denominator are all
averages over an annulus. These measurements allow us to estimate the
value of $g$ for the HST data. The central data point shows explicitly
that there is very little rotation in {\omegaCen} at the radii of our
central HST field. More generally, in the central 10 arcmin, the
rotation rate is approximately linear with radius, ${\cal V}_{\rm
pmt}/\Sigma_{\rm pmt} \approx R/750''$. Therefore, $g \approx
[1+(R/750'')^2]^{-1/2}$. These approximations are shown as solid lines
in the figure. This approximation implies that $0.995 \leq g \leq 1$
for the radii $R \leq 71.7''$ pertaining to the HST sample.

Given what is already known about proper motion rotation from the
vdV06 data, our estimates of $\Spmt$ should be quite accurate. This is
despite the fact that any solid-body part of the ${\cal V}_{\rm pmt}$
field cannot be measured directly in the HST data. In particular, the
value of $g$ is insufficient to explain the observed anisotropy
$\Spmt/\Spmr = 0.983 \pm 0.006$. Also, the data of vdV06 show that
$V_{\rm los}/\sigma_{\rm los}$ is about a factor two lower than ${\cal
V}_{\rm pmt}/\Sigma_{\rm pmt}$. Hence, the influence of rotation is
even more negligible for the line-of-sight measurements at $R=12''$ of
NGB08 (see discussion in Section~\ref{sss:groundlos}).


\newcommand{\figcapvsig}{
Observed ratios ${\cal V}_{\rm pmt}/\Sigma_{\rm pmt}$ (bottom; big
dots; magenta in the on-line version) and $g \equiv \Sigma_{\rm pmt} /
\Spmt$ (top; small dots; magenta), as function of projected radius $R$ from the
cluster center. The quantities in the nominator and denominator are
all averages over an annulus, and are based on the data of vdV06. The
solid lines indicate simple approximations to these quantities as
described in the text.  The dashed vertical line indicates the maximum
radius for which we use HST proper motions from Paper~I. The results
shown here indicate that our inability to determine ${\cal V}_{\rm
pmt}$ from the HST data does not prevent us from accurately
determining the RMS motion $\Spmt$ in the tangential proper motion
direction.\label{f:vsig}}

\begin{figure}[t]
\epsfxsize=\hsize
\centerline{\epsfbox{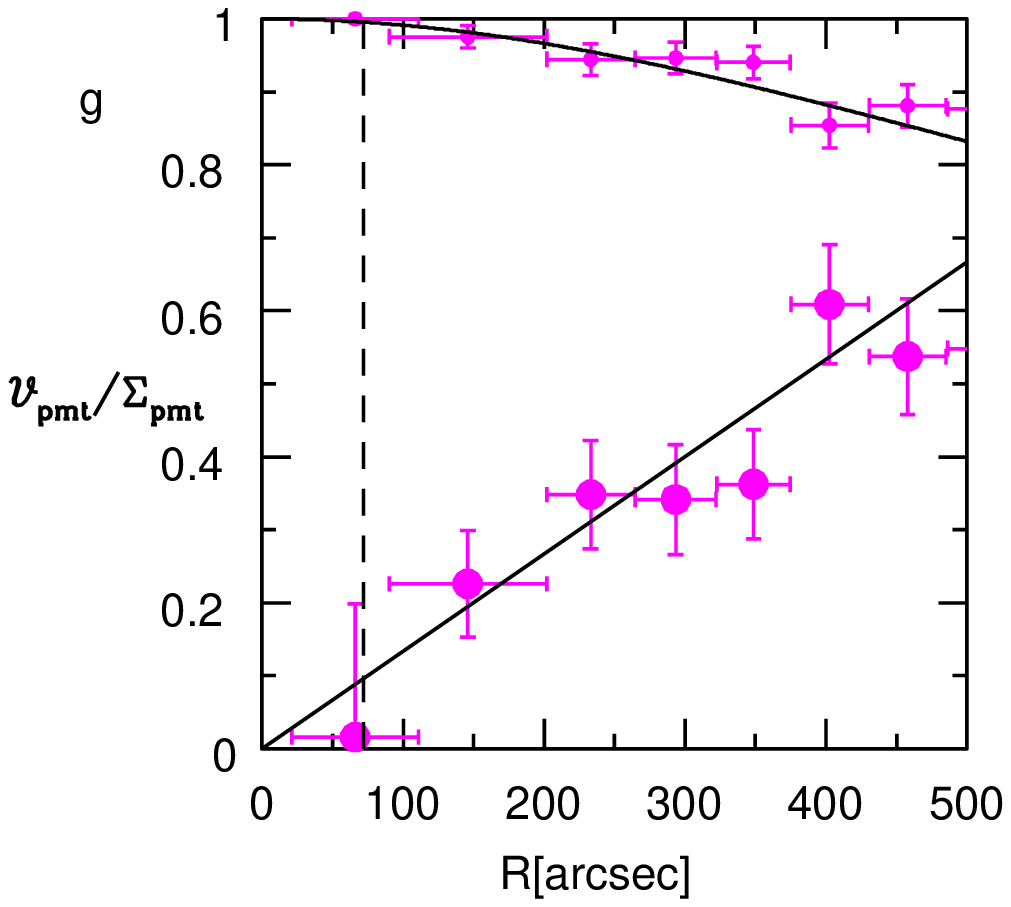}}
\figcaption{\figcapvsig}
\end{figure}



\newcommand{\figcapGH}{
Lowest order even Gauss-Hermite moments $h_4$ (top) and $h_6$ (bottom)
of the observed HST proper motion distributions as function of
projected radius $R$ from the cluster center. Different symbols
indicate distributions in the pmr (solid dots) and pmt (open circles)
directions. Data points in the two directions were slightly offset
horizontally, for visual clarity.  The Gauss-Hermite moments are not
far from zero, indicating that the proper motion distributions are
close to Gaussian. There are no systematic trends with radius. The
curves show predictions of three isotropic models, as discussed in
Section~\ref{ss:isoGH}. The long-dashed curves (red in the on-line
version) are for the isotropic core model with no dark mass. The solid
curves (blue) are for the isotropic cusp model with its best-fit IMBH
mass $\MBH = 1.8 \times 10^4 \Msun$. Both of these models match the
observed Gauss-Hermite moments to an average $|\Delta h_i| \lta 0.01$.
The short-dashed curve (green) is for the isotropic cusp model with no
dark mass. This model does not fit the observed Gauss-Hermite
moments.\label{f:GH}}

\begin{figure}[t]
\epsfxsize=\hsize
\centerline{\epsfbox{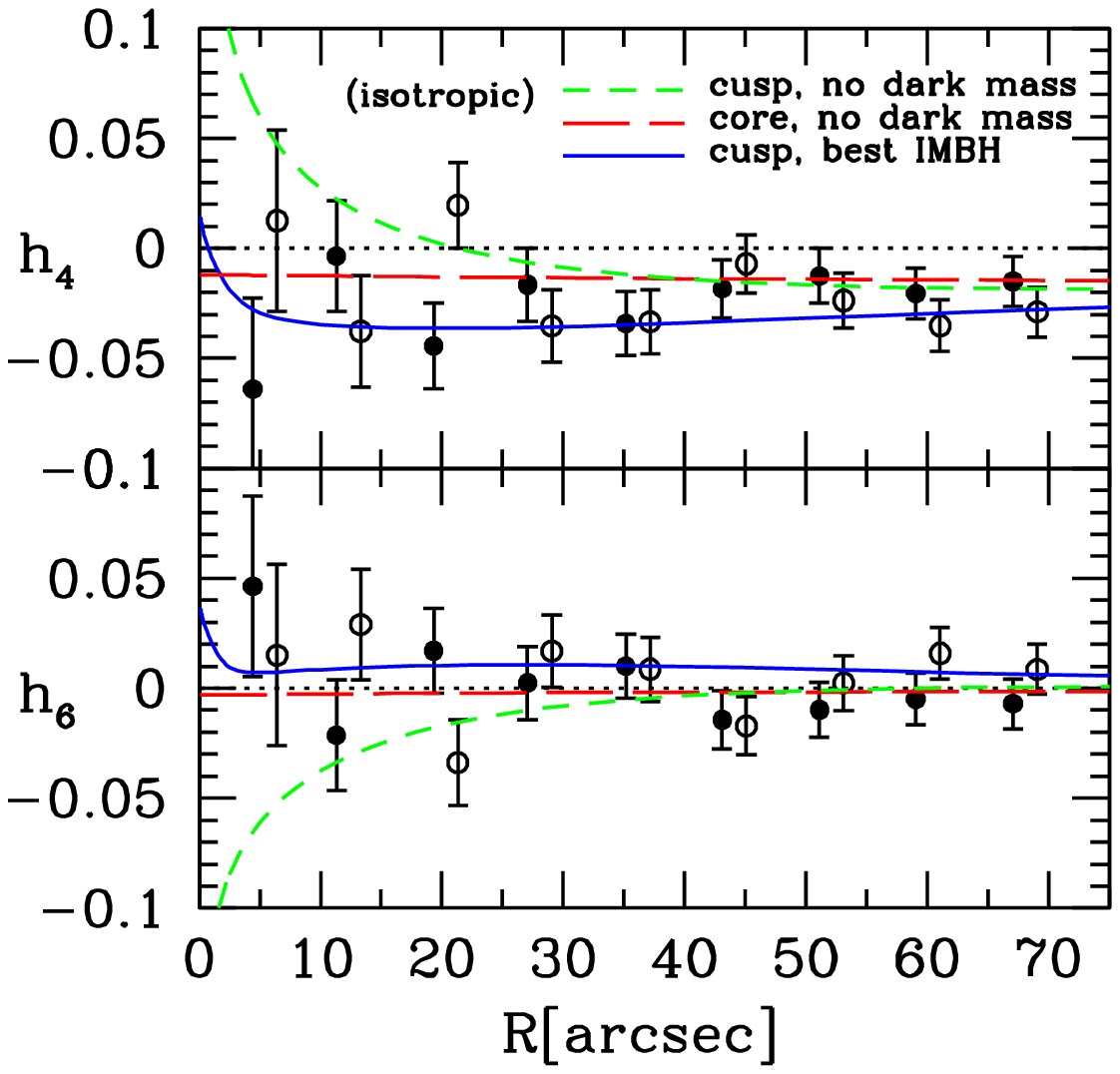}}
\figcaption{\figcapGH}
\end{figure}



\newcommand{\figcapanrat}{
Dynamical data-model comparisons for the ratio $\Spmt/\Spmr$ as
function of projected radius $R$ from the cluster center. The data and
the symbols are the same as in Figure~\ref{f:kin}c, and are the same
in both panels. (a; top panel) Anisotropic models with constant
$\beta$ as function of radius. The models shown have
$(\vnn/\vrr)^{1/2}$ ranging from $2/3$ to $3/2$, in equal logarithmic
steps. The corresponding $\beta$ values are indicated for some of the
curves. The horizontal line at $\Spmt/\Spmr = 1$ is for the isotropic
model. Tangentially anisotropic models have higher $\Spmt/\Spmr$ than
radially anisotropic models. The data are not well fit by models with
constant $\beta$. (b) Anisotropic model with $\beta(r)$ in
equation~(\ref{betadef}) chosen the optimize the fit to the data (blue
in the on-line version). The model has $\beta_0 = 0.13$,
$\beta_{\infty} = -0.53$ and $r_a = 729''$. The predictions in this
figure were calculated for a core model with no dark mass. However,
the predicted ratio ${\Spmt}/{\Spmr}(R)$ depends mostly on $\beta(r)$,
and is virtually indistinguishable when considering instead models
with a cusp or dark mass.\label{f:anrat}}

\begin{figure*}[t]
\epsfxsize=0.8\hsize
\centerline{\epsfbox{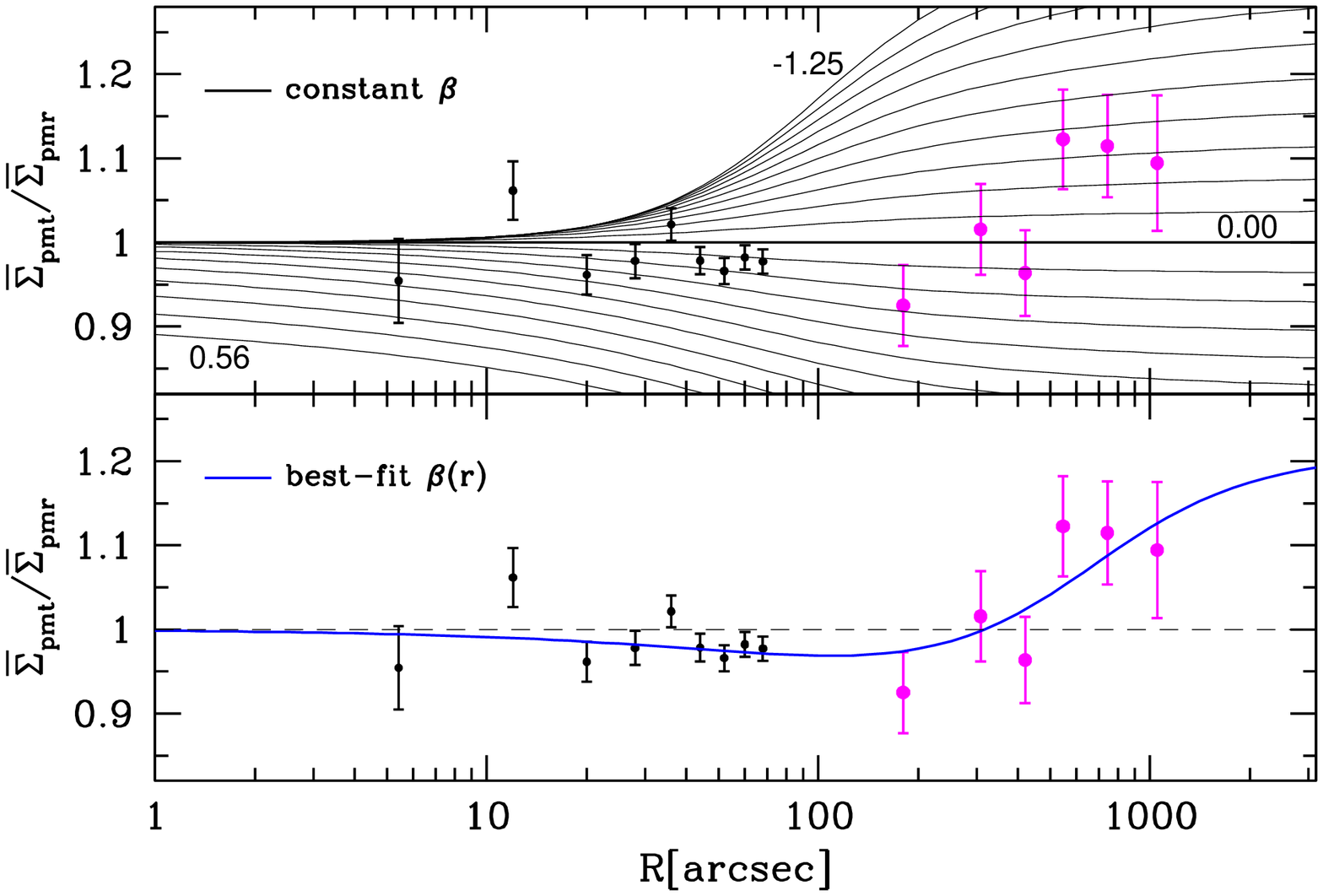}}
\figcaption{\figcapanrat}
\end{figure*}



\newcommand{\figcaprmsfit}{\footnotesize Dynamical data-model
comparisons for the RMS projected velocity as function of projected
radius $R$ from the cluster center. The data and the symbols are the
same as in Figure~\ref{f:kin}, and are the same in all panels. For
plotting purposes, the proper motions ${\Spm}$ in mas/year were
transformed to $\spm$ in km/s (solid symbols) using
equation~(\ref{pmunit}) with the best-fitting distance implied by the
models. The arrow in the bottom panel indicates how one of the two
data points from NGB08 moved due to our improved determination of the
cluster center in Paper~I. Curves are the predictions from various
models. Solid curves are for $\spm$ and short-dashed curves are for
$\slos$. For visual clarity, $\slos$ is not shown for $R \leq 10''$,
where no $\slos$ data are available. At larger radii, $\spm$ and
$\slos$ are quite similar, with differences becoming visual only at $R
\gta 1000''$. The curves that predict the lowest ${\bar \sigma}$ at
small radii (blue in the on-line version) are models with no dark
mass. The remaining curves (red) have some kind of dark mass in the
center. (a; top panel) anisotropic models in which $\beta(r)$ was
optimized to fit the data (see Figure~\ref{f:anrat}), based on the
core model for the projected intensity (see Figure~\ref{f:sbcen}). The
top curve has the best-fit IMBH of mass $\MBH = 4.1 \times 10^3
\Msun$. (b; middle panel) Anisotropic models similar to panel (a), but
now based on the cusp model for the projected intensity (see
Figure~\ref{f:sbcen}). The top curve has the best fit IMBH of mass
$\MBH = 8.7 \times 10^3 \Msun$. (c; bottom panel) Models similar to
panel (b), but now for an isotropic velocity distribution (which does
{\it not} fit the data for $\Spmt/\Spmr(R)$ in
Figure~\ref{f:anrat}). The quantities $\slos$ and $\spm$ are the same
for these isotropic models. The top three curves have different dark
masses. The solid curve is for the best-fitting IMBH mass $\MBH = 1.8
\times 10^4 \Msun$. The dotted curve is for the IMBH mass $\MBH = 4.0
\times 10^4 \Msun$ advocated by NGB08.  The long-dashed curve is for
the best fitting dark cluster, which has $M_{\rm dark} = 2.0 \times
10^4 \Msun$ and $a_{\rm dark} = 3.0''$. It follows from analysis of
the predictions in this figure, and in particular the top panel, that
the presence of an IMBH in {\omegaCen} is not required, although IMBHs
with masses $\MBH \lta 1.2 \times 10^4 \Msun$ cannot be ruled out at
$\sim 1\sigma$ confidence.\label{f:rmsfit}}

\begin{figure*}[t]
\epsfxsize=0.8\hsize
\centerline{\epsfbox{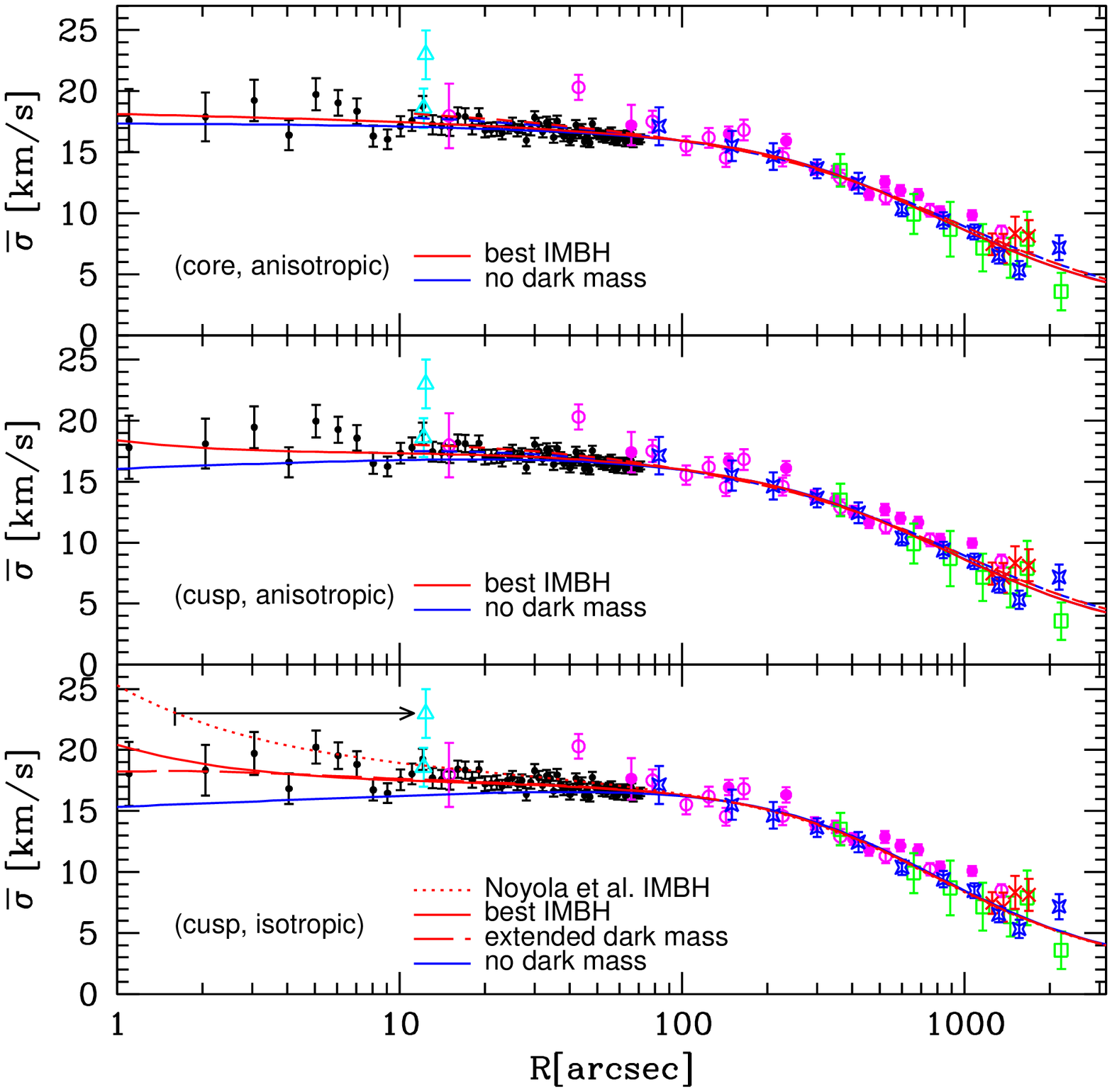}}
\figcaption{\figcaprmsfit}
\end{figure*}


\subsubsection{Gauss-Hermite Moments}
\label{sss:HSTGH}

To analyze the shapes of the HST proper motion distributions we binned
the data in concentric circular annuli as in Section~\ref{sss:HSTRMS}.
For each bin we created separate histograms of the proper motions in
the pmt and pmr directions. We used bins that correspond to $1 \kms$
for the canonical distance of {\omegaCen} ($D = 4.8 \kpc$; vdV06). For
each histogram we calculated the best-fitting Gaussian and the
corresponding Gauss-Hermite moments, as in van der Marel \etal
(2000). The odd moments are zero for distributions that are averaged
over circles. Figure~\ref{f:GH} shows the lowest order non-trivial
even Gauss-Hermite moments $h_4$ and $h_6$. These were calculated for
annuli that are $8''$ wide. The average values over all radii $R \leq
71.7''$ are $h_{4,{\rm pmr}} = -0.022
\pm 0.006$, $h_{4,{\rm pmt}} = -0.024 \pm 0.006$, $h_{6,{\rm pmr}} = -0.003 \pm
0.006$, $h_{6,{\rm pmt}} = 0.005 \pm 0.006$. The 8th and 10th order
moments were also calculated, but are not shown here. They were found
to be consistent with zero, like $h_6$. For all moments, the values
for the pmr and pmt directions were found to be consistent within the
errors. The only moment that is non-zero is $h_4$, with an average
value over the two proper motion directions of $h_4 = -0.023 \pm
0.004$. This is indicative of proper motion distributions that are
slightly more flat-topped than a Gaussian, and have slightly less
extended wings. We will compare this to model predictions in
Section~\ref{ss:isoGH}.

\section{Dynamical Data-Model Comparison}
\label{s:compare}

\subsection{Velocity Anisotropy}
\label{ss:aniso}

The observed profile of ${\Spmt}/{\Spmr}$ shows statistically
significant deviations from unity, both at small and at large radii
(see Figure~\ref{f:kin}c). It is therefore important to allow for
anisotropy in our models for {\omegaCen}.

To illustrate the relation between the model anisotropy function
$\beta(r)$ and the observed quantity ${\Spmt}/{\Spmr}(R)$, we show in
Figure~\ref{f:anrat}a the predicted ${\Spmt}/{\Spmr}(R)$ for models
with constant $\beta$ as a function of radius $r$. Models were
calculated with $(\vnn/\vrr)^{1/2}$ ranging from $2/3$ to $3/2$, in
equal logarithmic steps. At large radii, the proper motion anisotropy
is a good tracer of the intrinsic anisotropy, consistent with the
arguments presented in Section~\ref{sss:beta}. For example, at $R=10$
arcmin, $({\Spmt}/{\Spmr}) - 1 = \tau [(\vnn/\vrr)^{1/2} - 1]$, with
$\tau = 0.7$--0.9 for the models calculated here, and $\tau ~\approx
0.8$ for $\beta \approx 0$. At small radii, the behavior of
${\Spmt}/{\Spmr}$ is a more complicated function of
$\beta$. Tangentially anisotropic models ($\beta < 0$) all converge to
${\Spmt}/{\Spmr} = 1$ near the center, while radially anisotropic
models ($0 < \beta \leq 1$) predict ${\Spmt}/{\Spmr} < 1$.

It is evident from Figure~\ref{f:anrat}a that the observed
${\Spmt}/{\Spmr}(R)$ must tightly constrain the intrinsic
$\beta(r)$. We fitted the combined kinematical data from
Figure~\ref{f:kin} with models that spanned a grid in $\beta_0$,
$\beta_{\infty}$ and $r_a$. The parameter $\beta_0$ determines the
behavior of ${\Spmt}/{\Spmr}(R)$ near the center, $\beta_{\infty}$
determines the behavior at large radii, and $r_a$ is the transition
radius. The minimum $\chi^2$ was obtained for $\beta_0 = 0.13 \pm
0.02$, $\beta_{\infty} = -0.53 \pm 0.22$ and $\log r_a{\rm [arcsec]} =
2.86 \pm 0.12$. The error bars were obtained from the $\chi^2$
contours of the fit.  

The predicted ${\Spmt}/{\Spmr}(R)$ for the best-fit model is shown in
Figure~\ref{f:anrat}b. It provides an excellent fit. The predictions
shown in the figure were calculated for core models with no dark
mass. However, the predicted ratio ${\Spmt}/{\Spmr}(R)$ depends mostly
on $\beta(r)$, and is virtually indistinguishable when considering
instead models with a cusp or dark mass.

The error bars of the fit show that velocity anisotropy is detected at
high statistical significance in {\omegaCen}. The parameters $\beta_0$
and $\beta_{\infty}$ imply that the ratio $\vnn/\vrr$ transitions from
$0.935$ near the center to $1.235$ at large radii. These are of course
not necessarily the values at the very center and at infinity, since
our model is constrained by proper motion data only over the range
$1$--$1000$ arcsec. Either way, there is a transition from radial
anisotropy at small radii to tangential anisotropy at large radii. The
parameter $r_a$ indicates that the transition occurs at approximately
$12 \pm 3$ arcmin ($5 \pm 1$ core radii). This is broadly consistent
with the modeling results of vdV06. They found (their Section~9.2)
that {\omegaCen} is slightly radially anisotropic for $r \lta 10$
arcmin and that it becomes increasingly tangentially anisotropic
outside this region.

The centers of globular clusters are generally believed to have
isotropic velocity distributions because two-body relaxation tends to
isotropize the orbits. However, {\omegaCen} has a long half-mass
relaxation time of $\sim 10^{9.96 \pm 0.03}$ years (McLaughlin \& van
der Marel 2005). We showed in Paper~I that mass segregation in the
central region has not yet progressed to the point of energy
equipartition. Since full energy equipartition has not yet been
achieved, it is not surprising that complete velocity isotropy has not
yet been achieved either.

A sufficient number of stars is necessary to measure the proper motion
anisotropy with accuracy. Due to the finite number of stars at small
radii, the anisotropy is not as well constrained at radii $R \lta
10''$ as it is further out. However, because Omega Cen has a very
large core, most of the stars that are observed near the projected
center are not actually close to the center in three dimensions (see,
e.g., Section 6.1 of Paper I). One consequence of this is that the
projected kinematics predicted near the center are influenced
primarily by the anisotropy at larger radii. Changing the model
anisotropy in the central $10''$ even quite substantially (e.g., $\pm
30$\% in $\sigma_r/\sigma_t$) does not change the model predictions
for the HST proper motion kinematics by more than a fraction of the
error bars, even for the innermost data point. Therefore,
uncertainties in anisotropy near the center do not significantly
impact our predictions or conclusions.

\subsection{Cusp versus Core Models}
\label{ss:cuspcore}

Figures~\ref{f:rmsfit}a,b show the predicted and observed RMS
projected velocities for the best-fit $\beta(r)$. Each panel of this
figure shows the data for $\slos(R)$ and $\spm(R)$ together in the
same panel, with the observed proper motions $\Spm(R)$ transformed to
km/s using the best-fit distance implied by the model. For each model
there are separate curves predicted for $\slos(R)$ and $\spm(R)$
(dashed and solid curves, respectively), since these quantities are
not the same in an anisotropic model. However, the difference is
generally smaller than the observational error bars over the radial
range where both types of data are available. There is no
line-of-sight data available for $R \lta 10''$, so for visual clarity
we do not show the predicted $\slos(R)$ at those radii. At small radii
$\slos(R)$ exceeds $\spm(R)$ because of the model anisotropy, with
$\slos-\spm \approx 0.9 \kms$ at $R=10''$ and $\slos-\spm \approx 1.1
\kms$ at $R=1''$.

Figure~\ref{f:rmsfit}a shows the predictions for the core model fit to
the projected intensity, and Figure~\ref{f:rmsfit}b for the cusp model
fit. The curves that predict the lowest ${\bar \sigma}$ at small radii
(blue in the on-line version) are models with no dark mass. The core
and cusp models generally predict the same dynamics at large radii,
but differ slightly at small radii.

It is true in many analytical models with a cusped density profile and
no central dark mass that the projected RMS velocity decreases towards
the center (e.g., Tremaine \etal 1994). The same can be seen in
Figure~\ref{f:rmsfit}b for our model with $\gamma = 0.05$. Even though
the cusp is quite shallow in projection, this corresponds to a much
steeper three-dimensional luminosity density $j \propto r^{-1-\gamma}$
at asymptotically small radii. Similarly, the decrease in RMS
velocities towards the center is much stronger intrinsically in
three-dimensions than it is in projection. Either way, the observed
RMS velocities do not show a significant dip near the center. In the
absence of a dark mass, the core models therefore provide a better fit
to the data than the cusp models.

The best fits without a dark mass have $\chi^2 = 235.3$ and $249.2$
for the core and cusp models, respectively. There are 202 data points,
yielding $N_{\rm DF} = 197$ degrees of freedom (there are five free
parameters in the models when there is no dark mass). One expects
$\chi \approx N_{\rm DF} \pm \sqrt{2 N_{\rm DF}}$. The anisotropic
core model is therefore consistent with the data at the $1.9\sigma$
level, while the cusp model is consistent at the $2.6\sigma$ level.
These confidence levels assume that all data points in the fit are
statistically independent, which is somewhat of a simplification. It
is likely that samples from different line-of-sight velocity studies
have some stars in common. However, this cannot be fully explored here
since Scarpa \etal (2003) and Sollima \etal (2009) did not list
individually which stars they observed.

Much of the excess $E(\chi^2) \equiv \chi^2 - N_{\rm DF} > 0$ for the
fits is due to just two data points for $\slos$ that appear
anomalously high. The first ``discrepant'' data point is the
measurement $\slos = 23.0 \pm 2.0 \kms$ at $R = 12.4''$ by NGB08. The
second discrepant data point is the average $\slos = 19.9 \pm 1.0
\kms$ for an annulus centered at $R = 46.0''$, obtained from the
compiled data of vdV06. The former contributes $6.8$ to the $\chi^2$
of the fit for the core model, while the latter contributes
$11.3$. This adds up to $18.1$, which is almost half of $E(\chi^2) =
38.3$.

The discrepant NGB08 data point pertains to similar radius as the
other field that they observed, which is at $R = 12.1''$ from the
cluster center (see Paper~I). The measurement $\slos = 18.6
\pm 1.6 \kms$ for that latter field is more consistent with the other 
measurements in our combined data set from Section~\ref{s:kinematics}.
The two NGB08 measurements are inconsistent with each other at the
$1.7\sigma$ level, despite the fact that that they pertain to the same
radius, and despite the fact that the proper motions in these fields
are consistent with each other (Paper~I). We believe that this may be
due to underestimates in the NGB08 error bars, since they did not
explicitly include the contribution of shot noise (from the finite
number of stars) to their error bars. Similarly, the measurement in
the vdV06 annulus at $R = 46.0''$ appears inconsistent with other
measurements in that same data set. The weighted average of the two
adjacent annuli is lower and inconsistent with it at the $2.1\sigma$
level. This suggests that both discrepant measurements that contribute
disproportionately to $\chi^2$ may be spuriously high for unknown
reasons.

Evidently, there are some internal discrepancies in the datasets
themselves. Hence, the observed values of $E(\chi^2)$ should not be
taken as a sign of shortcomings in the models. In fact, the values of
$E(\chi^2)$ are remarkably low, given that we are fitting kinematical
data compiled from 10 different original sources. Potential
uncorrected systematics between different studies could easily have
induced much larger data-model discrepancies.

One clear result from Figures~\ref{f:rmsfit}a,b is that the data are
well fit by the models out to the largest available radii. This
contradicts arguments put forward by Scarpa \etal (2003) that the
large radii kinematics of {\omegaCen} may be inconsistent with
traditional theories of gravity. Our conclusion on this issue agrees
with that of McLaughlin \& Meylan (2003) and Sollima \etal (2009).


\newcommand{\figcapparab}{The $\chi^2$ of the model fits to the
kinematic data in Figure~\ref{f:kin}, as function of IMBH mass
$\MBH$. The resulting parabolae are, from bottom left to top right,
respectively, obtained for the following models: anisotropic models
with a core (black in the on-line version; see Figure~\ref{f:rmsfit}a
for actual model predictions); anisotropic models with a cusp (red;
see Figure~\ref{f:rmsfit}b); and isotropic models with a cusp (blue;
see Figure~\ref{f:rmsfit}c). Centered above the minimum of each
parabola is a triangle with an associated error bar that indicates the
best fit IMBH mass and its formal $1\sigma$ uncertainty. The
anisotropic core models provide the overall best fits to the
kinematics. The IMBH that optimizes the $\chi^2$ for those models is
not statistically significant. The small difference in $\chi^2$
(vertical height) between the parabolae for the anisotropic core and
cusp models is not a reason to prefer to core models over cusp models,
in particular because the latter actually provide a somewhat better
fit to the photometry (see discussion in the text).\label{f:parab}}

\begin{figure}[t]
\epsfxsize=\hsize
\centerline{\epsfbox{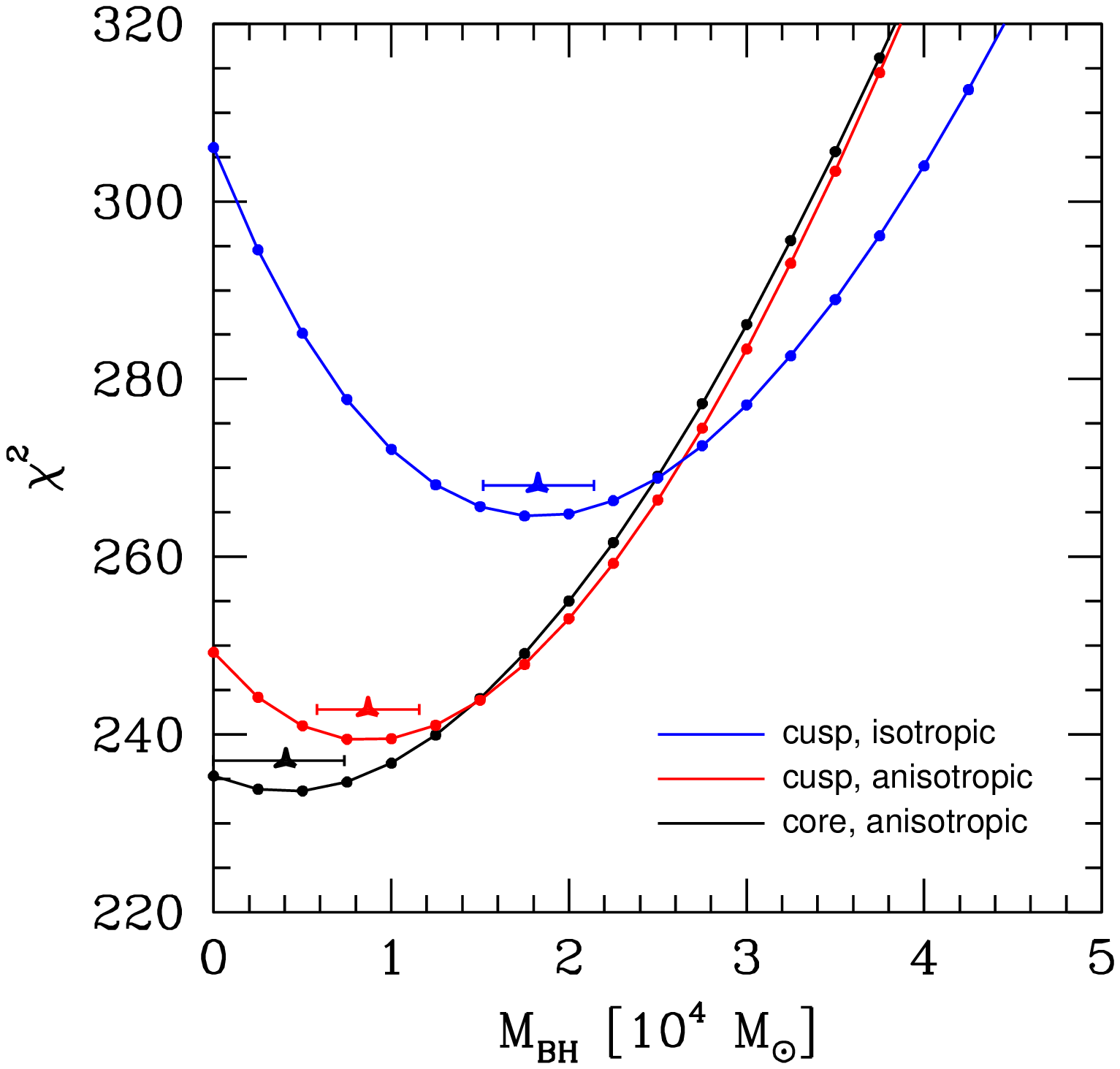}}
\figcaption{\figcapparab}
\end{figure}


\subsection{IMBH}
\label{ss:IMBH}

To test for the presence of an IMBH we have also constructed models
with $\MBH > 0$ (and the same $\beta(r)$ profile). Models with no IMBH
have one free parameter less than models with an IMBH. One would
therefore on average expect them to fit worse by $\Delta \chi^2 =
1$. Moreover, random variations can add another $\Delta \chi^2 = 1$
(at $1\sigma$ confidence), even when the no-IMBH model is
correct. Therefore, an IMBH is statistically significant in the model
only if it improves the fit over a no-IMBH model by $\Delta \chi^2
\gta 2$.

Figure~\ref{f:parab} shows the curves of $\chi^2(\MBH)$ that quantify
the quality of the fit. For the core model, the best-fit IMBH has mass
$\MBH = 4.1 \times 10^3 \Msun$. This improves the fit over the no-IMBH
model by only $\Delta \chi^2 = 1.8$. This is not statistically
significant. For the core model we can therefore determine only an
upper limit to the mass of a possible IMBH, namely $\MBH \leq 7.4
\times 10^3 \Msun$ at $1\sigma$ confidence. Figure~\ref{f:rmsfit}a compares 
the best-fit core models with and without an IMBH. The inclusion of
the IMBH makes very little difference to the predictions. The
difference for the innermost data point at $R=1.1''$ is only $0.8
\kms$, which is less than a quarter of the observational error bar at
that radius.

For the cusp model, the inclusion of an IMBH improves the fit by
$\Delta \chi^2 = 9.8$, which is significant at the $3.1\sigma$ level.
The best-fit IMBH mass is $\MBH = (8.7 \pm 2.9) \times 10^3
\Msun$ (with the error bar determined by the requirement $\Delta 
\chi^2 \leq 1$). Figure~\ref{f:rmsfit}b compares the best-fit cusp models 
with and without an IMBH. The inclusion of the IMBH provides a small
but visible improvement in the fit near the center. The reason that
the IMBH is more massive and significant in the cusp model is due to
the dip in RMS velocity predicted by these models when there is no
dark mass.

In assessing the meaning of formal error bars on $\MBH$, it is
important to keep in mind what one may call ``black hole
bias''. Statistical error bar estimates take into account only the
random Gaussian uncertainties in the data. They do not take into
account the systematic residuals that are inevitably present in any
astronomical data-model comparison.  In the present context these may
include oversimplifications in the model assumptions (e.g.,
sphericity, equilibrium, the adopted density parameterization, no mass
segregation among the luminous stars), the data reduction process
(e.g., inability to measure absolute proper motion rotation), or the
meaning of the data themselves (e.g., inability to separate orbital
motion from potential binary motion). By adding $\MBH$ as an extra
free parameter to the fit, the model gains in ability to fit these
systematic residuals that are not actually due to an IMBH. Because
masses are positive definite ($\MBH \geq 0$), this generally tends to
bias in the direction of invoking more mass then there actually is, as
compared to the opposite. Hence, $\MBH$ tends to be biased
high. Marginally significant non-zero $\MBH$ values with no tell-tale
signs of an IMBH (e.g., fast-moving stars or an $R^{-1/2}$ increase in
${\bar \sigma}$) may simply be indications of small systematic issues
in the data-model comparison unrelated to an IMBH.

The core model with no IMBH fits the kinematical data a little better
than the cusp model with an IMBH (see Figure~\ref{f:parab}). The
difference is $\Delta \chi^2 = 4.0$, which is nominally significant at
the $2.0\sigma$ level. However, as discussed in
Section~\ref{s:sbmodel}, the cusp model nominally provides a better
fit to the photometric data. The difference for a generalized nuker
fit is $\Delta \chi^2 = 6.8$, which is nominally significant at the
$2.6\sigma$ level. One could in principle construct a grand-total
$\Delta \chi^2$ that sums the photometric and kinematic
values. However, we have not done this here. The arguments about
whether a core or cusp model better fits the photometry are complex
(see Section~\ref{s:sbmodel}). They are not easily captured in a
single number without having to make arbitrary choices about, e.g.,
what radii to include in the comparison. Either way, when comparing
both the photometric and kinematic information for the core and cusp
models we feel that there is no sound statistical basis to prefer one
model over the other. In view of this, we conclude that the existing
data do not imply that an IMBH is present in {\omegaCen}. However,
IMBHs with masses $\MBH \lta 1.2 \times 10^4 \Msun$ cannot be ruled
out at $1\sigma$ confidence (or $\lta 1.8 \times 10^4 \Msun$ at
$3\sigma$ confidence).

\subsection{Distance and Mass-to-Light Ratio}
\label{ss:resdistups}

The best-fit distance and mass-to-light ratio depend only slightly on
the particular model we adopt. The core model without a dark mass has
$D = 4.70 \pm 0.06 \kpc$ and $\Upsilon_V = 2.64 \pm 0.03$ (here and
henceforth, mass-to-light ratios are given in $V$-band solar
units). The cusp model with an IMBH has $D = 4.75 \pm 0.06 \kpc$ and
$\Upsilon_V = 2.59 \pm 0.03$.

Our results for $D$ and $\Upsilon_V$ are entirely consistent with
those inferred by vdV06 from fitting only ground-based projected
intensity and kinematical data. They found that $D = 4.8 \pm 0.3 \kpc$
and $\Upsilon_V = 2.5 \pm 0.1$. Their modeling technique was
considerably more sophisticated than ours, allowing for axisymmetry,
inclination, ellipticity variations with radius, and arbitrary
profiles of rotation and velocity-dispersion anisotropy. The reason
that our error bars are smaller than theirs is likely a combination of
two effects. First, we have considerably more data at our disposal
than did vdV06, and in particular we have a factor $\sim 11$ more
stars with measured proper motions. This produces a significant
decrease in the random uncertainties. But second, in our simpler
models we cannot explore the full variety of phase-space distributions
that might fit the data with non-standard distances or mass-to-light
ratios. Therefore, we are not quantifying systematic errors as
rigorously as vdV06, and as a result our errors are artificially
lowered compared to theirs.

The fact that our $D$ and $\Upsilon_V$ agree with those of vdV06
indicates that there is no reason to mistrust that our spherical
models can address the central mass distribution of {\omegaCen} with
credible accuracy. This is further supported by the facts that both
the elongation (Geyer, Nelles, \& Hopp 1983) and the rotation rate
(Section~\ref{sss:HSTrot}) decrease towards the cluster center. To get
good results we did have to make sure to model RMS velocities (which
sum rotation velocities and velocity dispersions in quadrature) and
not merely velocity dispersions. Also, one should fit only quantities
that are properly averaged over annuli on the projected plane of the
sky. If these things are not done, then biased estimates are obtained
for $D$ and $\Upsilon_V$, as demonstrated in Appendix~C of vdV06.

\subsection{Isotropic Models}
\label{ss:iso}

Isotropic models predict that ${\Spmt}/{\Spmr} = 1$ at all radii, so
they are not generally appropriate for {\omegaCen}.  However,
isotropic models have the advantage that the full velocity
distributions $\cL_{\rm iso}(v,R)$ can be calculated as described in
Section~\ref{ss:vp}. Also, isotropic models are often used for
globular clusters, so it is useful to understand how the predictions
of such models deviate from more general anisotropic models.

Figure~\ref{f:parab} shows the curve of $\chi^2(\MBH)$ for isotropic
models with an IMBH. The models fit significantly worse than the
anisotropic models, as shown by the higher $\chi^2$. Within the realm
of isotropic models, the best fit is provided by $\MBH = (1.8
\pm 0.3) \times 10^4 \Msun$. This model has $D = 4.81 \pm 0.06 \kpc$
and $\Upsilon_V = 2.61 \pm 0.03$.  Figure~\ref{f:rmsfit}c shows the
data-model comparison for the isotropic model with this IMBH. We also
show the predictions for $\MBH = 0$, as well as for the value $\MBH =
4.0 \times 10^4 \Msun$ advocated by NGB08. 

The NGB08 IMBH mass was based on their measurement of $\slos = 23.0
\pm 2.0 \kms$ at a position they believed to be the cluster
center. This high IMBH mass is now clearly ruled out, even for an
isotropic model, for two reasons. First, Paper~I showed that this
measurement actually applies to a larger distance $R = 12.4''$ from
the cluster center. The arrow in Figure~\ref{f:rmsfit}c demonstrates
how this moved the corresponding line-of-sight velocity
measurement. And second, the high measured RMS velocity was not
confirmed by proper motion measurements, either at the same position
or on the actual cluster center. Note in Figure~\ref{f:rmsfit}c that
the isotropic model with the NGB08 IMBH does appear to match well the
data at $R=5$--$7''$. However, it overpredicts the data at both $R
\leq 4''$ and $R=8$--$10''$. Since all data points were derived in the
same manner and are equally valid, there is no reason attribute more
weight to the data at $R=5$--$7''$. In particular, the data at $R \leq
4''$ contain most of the information on the very central mass
distribution, and they clearly rule out an IMBH mass as high as
advocated by NGB08 (independent of the model anisotropy near the
center).

It is evident from Figure~\ref{f:parab} that isotropic models require
a higher IMBH mass than the best-fit anisotropic models. This is not
generally true for isotropic models, but is a specific consequence of
the fact that {\omegaCen} is radially anisotropic near the center and
tangentially anisotropic at large radii. This affects the general
gradient in RMS velocity from small to large radii (see e.g.,
Figures~10 and~11 of van der Marel~1994). The anisotropic model
without a dark mass has a steeper gradient towards the center than the
corresponding isotropic model. The anisotropic model therefore
requires less dark mass to fit the observed gradient. This indicates
that one should be careful in application of isotropic models to
studies of IMBHs. Depending on the exact anisotropy of the cluster,
the assumption of isotropy can lead to either an overestimated or an
underestimated IMBH mass, or it can require an IMBH when none is
present.


\newcommand{\figcapjeansbest}{
Contours of $\Delta \chi^2$ in the two-dimensional plane of $(a_{\rm
dark}, M_{\rm dark})$ for the fit of isotropic cusp models to the data
in Figure~\ref{f:kin}. The distance $D$ and mass-to-light ratio
$\Upsilon_V$ were optimized separately at each $(a_{\rm dark}, M_{\rm
dark})$ for which a model was calculated (indicated with small
dots). The model for which the minimum $\chi^2$ was obtained is
indicated by the big dot. The first three contours (solid) around the
minimum correspond to $1\sigma$, $2\sigma$, and $3\sigma$ (bold)
confidence contours, respectively. Each subsequent contour (dashed)
increases $\Delta \chi^2$ by a constant factor. Models with a central
dark points mass (an IMBH; i.e., $a_{\rm dark} = 0$) are shown on the
vertical axis on the left. Models with no central dark mass lie on the
horizontal axis on the bottom. If there is a dark mass in {\omegaCen}
(as must be the case in isotropic models, but these do {\it not} fit
the data for $\Spmt/\Spmr(R)$ in Figure~\ref{f:anrat}) then its extent
must be $\lta 7''$ at $1\sigma$ confidence.\label{f:jeansbest}}

\begin{figure}[t]
\epsfxsize=\hsize
\centerline{\epsfbox{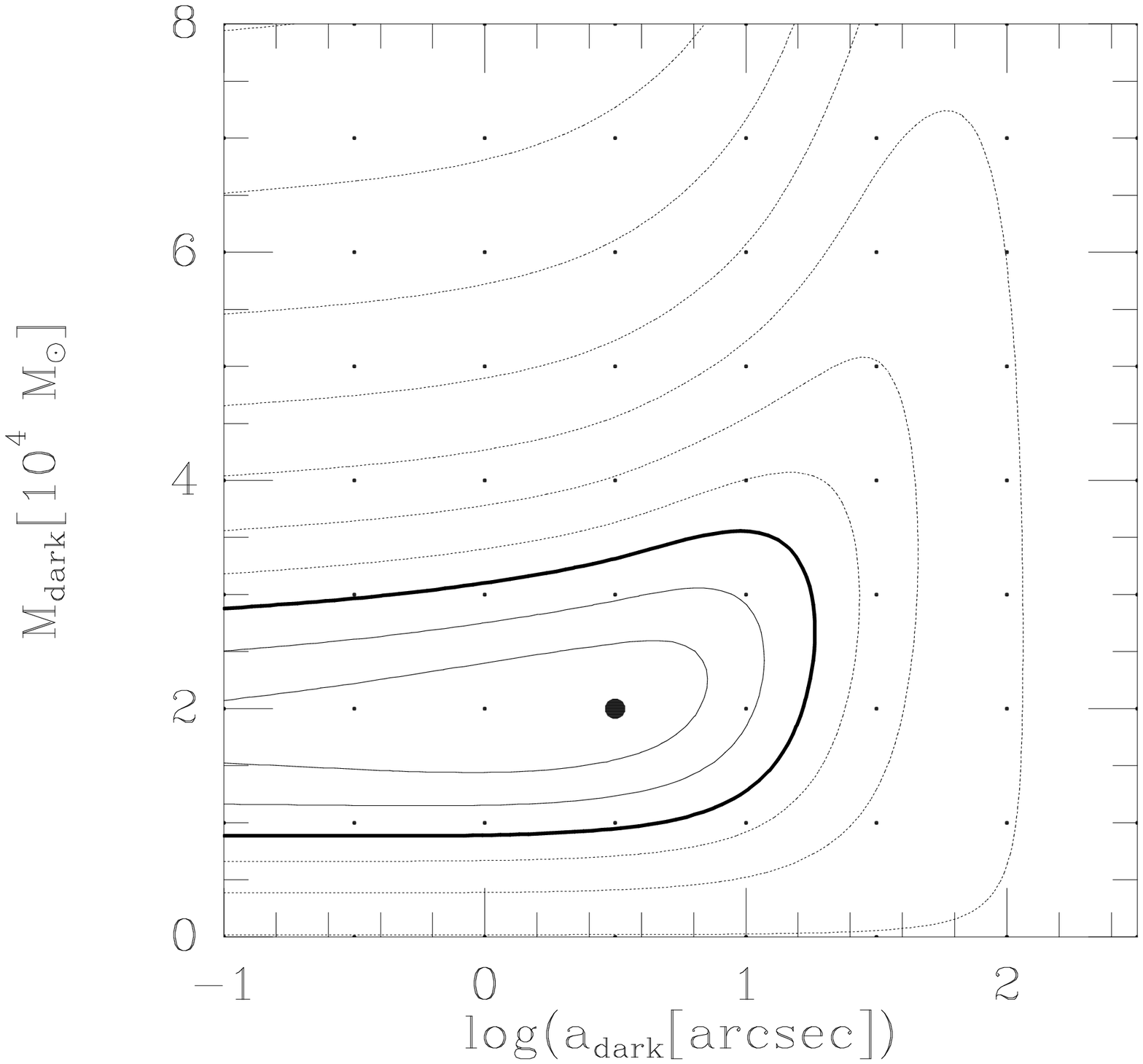}}
\figcaption{\figcapjeansbest}
\end{figure}



\newcommand{\figcapVP}{
Symmetrized histograms of the observed proper motion velocities for
apertures of radii $R=3''$ (bold; red in the on-line version) and
$R=10''$ (thin; blue), respectively. Proper motions in both the pmr
and pmt directions are included. They were transformed to physical
units of km/s using the best-fit model distance. Typical error bars
are shown on the left; they were offset horizontally from the
histograms for visual clarity.  The predictions for the isotropic cusp
model with the best fit IMBH mass $\MBH = 1.8 \times 10^4 \Msun$ are
over-plotted as curves (solid for the $R=3''$ aperture, and dashed for
the $R=10''$ aperture). The inset on the right shows a blow-up of the
profile wings on an expanded vertical scale. The models predict
broader wings than observed, but the difference is not statistically
significant (as discussed in the text). This is because the total area
under the model curves beyond $60 \kms$ corresponds to less than 1
predicted star.\label{f:VP}}

\begin{figure}[t]
\epsfxsize=\hsize
\centerline{\epsfbox{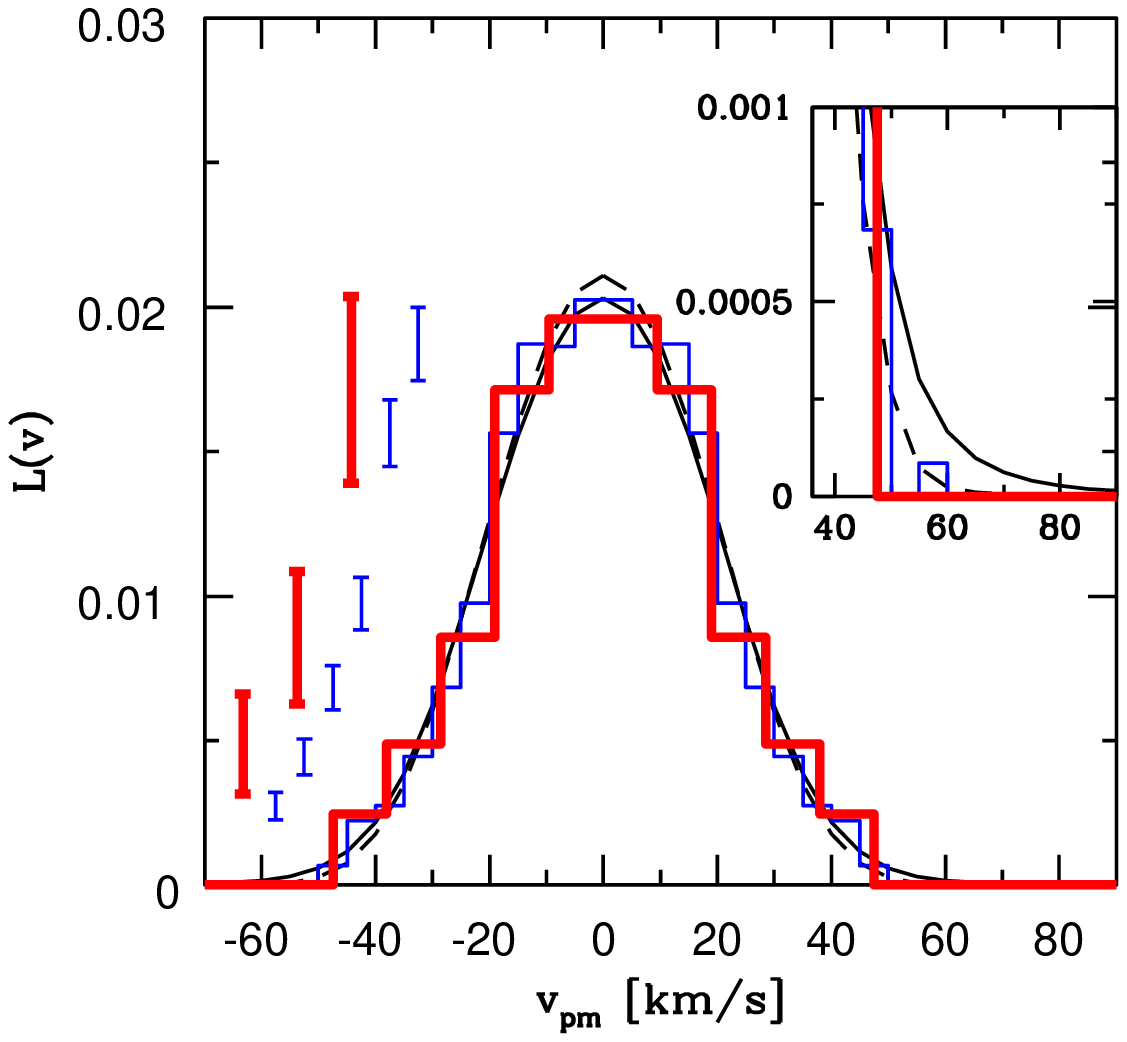}}
\figcaption{\figcapVP}
\end{figure}


\subsection{Dark cluster}
\label{ss:darkclus}

When a dark mass is clearly required by the data, as in the case of
isotropic models with a cusp, one can ask whether an extended dark
cluster may provide a better fit than an
IMBH. Figure~\ref{f:jeansbest} shows a two-dimensional contour plot of
$\Delta \chi^2$ in the parameter space of $(a_{\rm dark}, M_{\rm
dark})$ for the isotropic cusp model. The best fit is obtained for
$M_{\rm dark} = 2.0 \times 10^4 \Msun$ and $a_{\rm dark} = 3''$. The
predictions of this model are shown in Figure~\ref{f:rmsfit}c. It has
a somewhat shallower increase in RMS velocity towards the center than
the IMBH model. The dark-cluster model makes an improvement of only
$\Delta \chi^2 = 1.7$ over the IMBH model, while increasing the number
of free parameters by one. This is not statistically significant.
Therefore, one can interpret Figure~\ref{f:jeansbest} to mean that if
there is a dark mass in {\omegaCen}, then its extent must be $\lta
7''$ at $1\sigma$ confidence.

The central density of the best-fitting dark cluster is $\rho_{\rm
0,dark} \equiv M_{\rm dark} / (4 \pi a_{\rm dark}^3 / 3)$, according
to equation~(\ref{rhodark}). Using equation~(\ref{radscale}) to
transform angular units to physical units, this yields for the core
model $\rho_{\rm 0,dark} = 1.4 \times 10^{7} \Msun \pc^{-3}$. The
central density of luminous matter in this model is only $\rho_{\rm
0,lum} = 3.3 \times 10^3 \Msun \pc^{-3}$. This implies a central $M/L
= \Upsilon_V [1 + (\rho_{\rm 0,dark}/\rho_{\rm 0,lum})] = 1.1 \times
10^4$. This applies to an isotropic model with $M_{\rm dark} = 2.0
\times 10^4 \Msun$. For an anisotropic model with a lower $M_{\rm
dark}$, the required $M/L$ scales approximately linearly with $M_{\rm
dark}$. The (maximum) size $a_{\rm dark}$ of the dark mass fit is
relatively insensitive to the exact anisotropy. Either way, if
{\omegaCen} does indeed have a dark central cluster, this would be
indicative of quite an extreme amount of mass segregation or a very
top-heavy initial mass function. If a dark mass is invoked to explain
the data, then an IMBH may be easier to explain than such a
concentrated dark cluster.

\subsection{Velocity Distribution Shapes}
\label{ss:isoGH}

For our isotropic models we calculated the projected velocity
distributions $\cL_{\rm iso}(v,R)$ as described in
Section~\ref{ss:vp}, and from this the Gauss-Hermite moments
$h_{i,{\rm iso}}(R)$. We did not take the effect of observational
proper motion uncertainties into account in the models (essentially a
convolution of the model predictions with the error
distribution). This has negligible effect on the Gauss-Hermite moments
because the observational errors are much less than the intrinsic
cluster dispersion. The curves over-plotted in Figure~\ref{f:GH} show
the model predictions for three models: the isotropic core model with
no dark mass, the isotropic cusp model with its best-fit $\MBH = 1.8
\times 10^4 \Msun$, and the isotropic cusp model with no dark mass. 

The core model with no dark mass and the cusp model with an IMBH both
fit the observed Gauss-Hermite moments within the error bars. Over the
range $5'' \lta R \leq 71.7''$, the predicted Gauss-Hermite moments
for these models are approximately constant. For the core model
without dark mass they are $h_{4,{\rm iso}}(R) \approx -0.013$ and
$h_{6,{\rm iso}}(R) \approx -0.002$, and for the cusp model with an
IMBH they are $h_{4,{\rm iso}}(R) \approx -0.034$ and $h_{6,{\rm
iso}}(R) \approx 0.009$. These averages do deviate in statistically
significant manner from the observed averages, $h_4 = -0.023 \pm
0.004$ and $h_6 = 0.001 \pm 0.004$ (see Section~\ref{sss:HSTGH}).
This is not surprising, given that we have already established that
{\omegaCen} does not have an isotropic velocity distribution, not even
near its center. However, the differences between the predicted and
the average observed Gauss-Hermite moments are quite small $|\Delta
h_i| \lta 0.01$. For comparison, models that range in anisotropy from
fully tangential to fully radial can easily span the range $\Delta h_i
\approx \pm 0.2$ (e.g., van der Marel \& Franx 1993). The 
measured Gauss-Hermite moments are therefore consistent with the fact
that the velocity distribution is not far from isotropic near the
center of {\omegaCen}.

The cusp model without dark mass does {\it not } fit the observed
Gauss-Hermite moments. As discussed in Section~\ref{ss:cuspcore}, this
model has a strong decrease in its intrinsic (unprojected) RMS
velocity towards the center. The decrease in the projected RMS
velocity is relatively mild, as shown in Figure~\ref{f:rmsfit}c.
However, there is a strong gradient in the RMS velocity along the line
of sight. This leads to strongly non-Gaussian velocity distributions.
In a central aperture of $R=8''$ radius, as used for the innermost
observational data point in Figure~\ref{f:GH}, the predictions are
$h_{4,{\rm iso}} = 0.063$ and $h_{6,{\rm iso}} = -0.063$. This
deviates for each order by $\sim 3\sigma$ from the observed values
$h_4 = -0.026 \pm 0.029$ and $h_6 = 0.031 \pm 0.029$ (averaged over
the pmr and pmt directions).  While this result was derived for
isotropic models, the large predicted deviations from Gaussian
velocity distributions are likely generic, and would exist in
anisotropic models as well. So if {\omegaCen} has a central cusp in
its projected intensity profile, then the observed velocity
distributions rule out models without a dark mass. Of course,
Figure~\ref{f:rmsfit}b,c show that such models also do not provide
good fits to the projected profiles of RMS velocity.

Figure~\ref{f:GH} shows that an isotropic model {\it with} an IMBH
predicts increased $h_4$ and $h_6$ in the central $R \lta 5''$,
corresponding to profiles with broader wings than a Gaussian. To do a
more detailed data-model comparison for the wings we show in
Figure~\ref{f:VP} the histograms of the observed proper motions for
apertures of radii $R=3''$ and $R=10''$. These apertures contain $N =
43$ and $N = 585$ stars with well-measured proper motions,
respectively. The proper motion coordinates in the pmr and pmt
direction were both included in each histogram, as appropriate for an
isotropic velocity distribution. The histograms were symmetrized to
decrease the uncertainties. The predictions for the isotropic cusp
model with $\MBH = 1.8 \times 10^4 \Msun$ are over-plotted as
curves. The predicted velocity distribution for the $R=3''$ aperture
has more extended wings than the observed histogram.  The wings are
much reduced in the predicted velocity distribution for the $R=10''$
aperture. This is because none of the added stars at $3'' < R \leq
10''$ are close to the black hole, so they do not move at unusually
high velocities.

The observed histograms do not extend beyond $60 \kms$. The maximum
observed value of either $|v_{\rm pmt}|$ or $|v_{\rm pmr}|$ inside
$R=10''$ is $59.4 \kms$. No stars were rejected as non cluster members
inside this radius because of their fast motion (see
Section~\ref{sss:HSTsample}). To assess whether these observations are
statistically consistent with the wings of the predicted
distributions, let $f$ be the fraction of the normalized model
distribution at $|v| > 60 \kms$. According to the models, this is $f =
3.8 \times 10^{-3}$ for the $R=3''$ aperture and $f = 3.9 \times
10^{-4}$ for the $R=10''$ aperture.  The expectation value for the
number of stars with a velocity component at $|v| > 60 \kms$ is $E = 2
N f$. The probability that all of the $2N$ measured velocities have
$|v| \leq 60 \kms$ is $P = (1-f)^{2N}$. This yields for the $R=3''$
aperture that $E = 0.32$ and $P=0.72$, while for the $R=10''$ aperture
$E = 0.46$ and $P=0.63$.

Since less than a single fast-moving star was predicted, the fact none
were observed is not statistically inconsistent with the models. This
does not change when we use larger apertures, since the models do not
predict any fast-moving stars outside $R=10''$. It also does not
change by using smaller apertures, since in that case the sample of
observed stars becomes very small. Therefore, the absence of
fast-moving stars in the data cannot be used to independently rule out
the presence of an IMBH with mass $\MBH = 1.8 \times 10^4 \Msun$. To
increase the predicted number of fast-moving stars one would need to
increase the number of stars with measured proper motions, i.e.,
extend the kinematical observations to fainter magnitudes. We have
experimented with looser cuts on our proper motion catalog. This
allows us to increase the sample size by a factor $\sim 2$ (see
Paper~I). In this extended catalog there is no star with a proper
motion coordinate in excess of $65 \kms$ within $R = 10''$. This
provides somewhat tighter constraints on the presence of an IMBH, with
the quoted values of $P$ roughly decreasing to their squared
values. Even then, it is still not possible to rule out an IMBH of
mass $\MBH = 1.8 \times 10^4 \Msun$ with reasonable confidence.

Alternatively, $E$ increases and $P$ decreases when the BH mass is
increased. For the IMBH of mass $\MBH = 4.0 \times 10^4 \Msun$
suggested by NGB08, and sticking as before with our high-quality
proper motion catalog of Section~\ref{sss:HSTsample}, we obtain
$E=2.9$ and $P=0.05$ for the aperture with $R=10''$. This implies that
the absence of fast-moving stars rules out all IMBHs with $\MBH \gta
4.0 \times 10^4 \Msun$ at 95\% confidence. This result is not very
competitive for {\omegaCen}, since we already know from modeling of
the RMS velocities that any IMBH must have a mass at least $3.3$ times
lower than this anyway. However, this method has the advantage that it
does not require any accurate measurement of RMS velocities. It may
therefore be useful for other clusters, in cases when the
observational errors may be too large for accurate measurement of the
RMS velocity profile.


\newcommand{\tablecontentsdemo}{
\tablecaption{Globular clusters with IMBH constraints\label{t:demo}} 
\tablehead{
\colhead{cluster} & 
\colhead{$D$} &
\colhead{$V_0$} &
\colhead{$M/L_V$} &
\colhead{$M_{\rm tot}$} & 
\colhead{$\sigma_{\rm e}$} &
\colhead{$M_{\rm BH}$[Tre]} &
\colhead{$M_{\rm BH}$[obs]} &
\colhead{data} &
\colhead{$M_{\rm BH}$[obs] / $M_{\rm tot}$} \\
\colhead{} & \colhead{(kpc)} & \colhead{(mag)} & 
\colhead{($\odot$)} & \colhead{($\Msun$)} & \colhead{(km/s)} &
\colhead{($\Msun$)} & \colhead{($\Msun$)} & & \\
\colhead{(1)} & \colhead{(2)} & \colhead{(3)} & \colhead{(4)} & \colhead{(5)} &
\colhead{(6)} & \colhead{(7)} & \colhead{(8)} & \colhead{(9)} & 
\colhead{(10)} \\ }
\startdata
47 Tuc & 4.5$^b$ & 3.91$^b$ & 2.4$^b$ & $1.1 \times 10^6$ & 9.8$^k$ &
$7.3 \times 10^2$ & $<1.5 \times 10^3$$^g$ & discrete PM & $<0.13$\%
\\ 
Omega Cen & 4.7$^a$ & 3.14$^b$ & 2.6$^a$ & $2.8 \times 10^6$ &
15.7$^j$ & $4.8 \times 10^3$ & $<1.2 \times 10^4$$^a$ & discrete
PM+LOS & $<0.43$\% \\ 
M15 & 10.3$^l$ & 5.89$^e$ & 1.6$^c$ & $6.2
\times 10^5$ & 12.1$^c$ & $1.7 \times 10^3$ & $<4.4 \times 10^3$$^h$ &
discrete PM+LOS & $<0.71$\% \\ 
NGC 2298 & 10.7$^b$ & 9.36$^b$ &
1.9$^b$ & $3.4 \times 10^4$ & 2.4$^k$ & $2.6 \times 10^0$ & $<3.4
\times 10^2$$^f$ & mass segregation & $<1.0$\% \\ 
G1 & 770$^i$
&13.49$^i$ & 2.8$^d$ & $5.7 \times 10^6$ & 25.1$^i$ & $3.2 \times
10^4$ & $(1.8 \pm 0.5) \times 10^4$$^d$ & unresolved LOS &
$\!\!\!\!\!\!\!\!\!\!\!\!\!\!(0.32 \pm 0.09)$\% \\
\enddata}

\newcommand{\tablecommdemo}{Column~(1) lists the cluster name.
Column~(2) lists the cluster distance $D$. Column~(3) lists the
extinction-corrected total apparent magnitude $V_0$. Column~(4) lists
the $V$-band mass-to-light ratio $M/L_V$ in units of
$\Msun/\Lsun_V$. Column~(5) lists the total cluster mass $M_{\rm tot}$
derived from columns~(2)--(4). Column~(6) lists the effective velocity
dispersion $\sigma_{\rm e}$. Column~(7) lists the value of $M_{\rm
BH}$ predicted by the $M_{\rm BH}$ vs.~$\sigma_e$ relationship for
galaxies of Tremaine \etal (2002), using the value of $\sigma_{\rm e}$
in column~(6). The uncertainty in the predicted $M_{\rm BH}$ due to
the uncertainties in the relation itself is a factor $\sim 1.5$. There
is also intrinsic scatter about the relationship of a factor $\sim
2$. Column~(8) lists the observational constraint on the mass $M_{\rm
BH}$ of any IMBH. Column~(9) lists the type of data on which the
observational constraint in column~(8) is based. This can be either
discrete velocities in the proper motion (PM) and/or line-of-sight
(LOS) directions; unresolved observations of line-of-sight kinematics;
or measurements of the amount of mass segregation, which can be
compared to model predictions. Column~(10) lists the observational
constraint on the ratio $M_{\rm BH}/M_{\rm tot}$. Sources of the
numbers in the table are: $^a$ This paper. $^b$ McLaughlin \& van der
Marel (2005). $^c$ Gerssen \etal (2002). $^d$ Gebhardt \etal
(2005). $^e$ Harris (1996). $^f$ Pasquato \etal (2009). $^g$
McLaughlin \etal (2006). $^h$ Gerssen \etal (2003). $^i$ Meylan \etal
(2001). $^j$ Line-of-sight velocity dispersion in an aperture of size
equal to the effective radius, calculated from the models in this
paper; this is 85\% of the central value. $^k$ The central model
velocity dispersion value in McLaughlin \& van der Marel (2005),
multiplied by the same 85\% correction derived in footnote $j$. $^l$
van den Bosch \etal (2006).}

\begin{deluxetable*}{lrrrrrrrlr}
\tabletypesize{\tiny}
\tablecontentsdemo
\tablecomments{\small\tablecommdemo}
\end{deluxetable*}


\subsection{Brownian Motion}
\label{ss:brown}

We have assumed in the modeling that any IMBH must reside at the
cluster center. In reality, an IMBH would exhibit Brownian motion as a
result of its energy equipartition with surrounding stars. Chatterjee,
Hernquist \& Loeb (2002) derived for the one-dimensional RMS offset
from the cluster center that
\begin{equation}
  x_{\rm RMS} = (2 {\tilde m} / 9 \MBH)^{1/2} \> r_0 .
\end{equation}
This was derived for an analytical Plummer model with core radius
$r_0$, which provides a reasonable description of (non core-collapsed)
globular clusters. Chatterjee \etal considered single-mass models with
stars of mass ${\tilde m}$. However, their results should be valid for
multi-mass models as well, provided that the quantity ${\tilde m}
\equiv \langle m^2 \rangle / \langle m \rangle$ is defined appropriately in
terms of averages over the stellar mass function (Merritt, Berczik \&
Laun 2007).

We take for $r_0$ the King core radius of $141''$ (McLaughlin \& van
der Marel 2005) and for $\MBH$ the upper limit of $1.2 \times 10^4
\Msun$ derived in Section~\ref{ss:IMBH}. For a Salpeter mass 
function from $0.1$--$10 \Msun$, ${\tilde m} = 1.27 \Msun$. We then
obtain $x_{\rm RMS} = 0.69''$. This is significantly less than the
$1.55''$ radius of the smallest aperture used in the kinematical
analysis (see Section~\ref{sss:HSTRMS}). Therefore, the expected
Brownian motion of an IMBH should not affect significantly the
analysis that we have presented here.

IMBHs with larger masses than those considered here, placed at
considerable offsets from the cluster center (consistent with their
expected Brownian motion), might have gone unnoticed in our
analysis. However, there is no independent motivation to consider such
a situation. In Paper~I we showed that the kinematical center of stars
in {\omegaCen} agrees with the star count center to within its $\sim
2''$ uncertainty.

\section{Globular Cluster IMBH Demographics}
\label{s:demographics}

The self-gravity of a cluster gives it a negative heat capacity that
makes it vulnerable to the so-called ``gravothermal
catastrophe''. This makes the core collapse on a timescale
proportional to the half-mass relaxation time. This is normally halted
by formation and/or heating of binaries. However, the most massive
stars in a cluster tend to undergo core collapse more or less
independently of the other cluster stars on a timescale that is less
than the core collapse time for the cluster as a whole. If this
happens faster than the main sequence evolution time of these stars
(i.e., before they become compact remnants), then they will undergo
physical collisions that lead to the formation of a single Very
Massive Star (VMS). Such a VMS could plausibly evolve into an
IMBH. 

Portegies Zwart \& McMillan (2002) proposed the aforementioned
scenario and performed $N$-body simulations to show that a VMS might
grow to $M_{\rm VMS} / M_{\rm tot} \approx 0.1$\%, with $M_{\rm tot}$
being the total cluster mass. Gurkan \etal (2004) and Freitag \etal
(2006) used Monte-Carlo simulations with prescriptions and models for
stellar collisions and found that the resulting VMS could even grow up
to $M_{\rm VMS} / M_{\rm tot} \approx 1$\%. However, it is not clear
what the resulting IMBH mass might be. The evolution of a VMS may
yield significant mass loss before the final collapse. Also, a VMS
will only form through this scenario in clusters with very short
initial half-mass relaxation times, $\lta 30$ Myr. Some star clusters
(e.g., the Arches cluster) are known to have such short relaxation
times. However, most globular clusters have much longer current
relaxation times, although it is not straightforward to estimate what
their relaxation time at birth may have been (Ardi \etal 2008).

Alternative scenarios for IMBH formation in globular clusters have
been proposed as well (see e.g. the review in van der Marel 2004).
This includes the merging of stellar-mass black holes. O'Leary \etal
(2006) found that this may lead to growth up to $M_{\rm IMBH} / M_{\rm
tot} \approx 0.1$\%. Whether this indeed happens depends on the
typical BH merger recoil speed. Significant recoils will kick black
holes out of the cluster before they have a chance to grow.

Overall, the predictions for IMBH formation in globular clusters have
significant uncertainties. Observations will therefore need to be the
primary guide for the study of IMBH demographics in globular
clusters. For {\omegaCen} one may combine the distance and
mass-to-light ratio derived in Section~\ref{ss:resdistups} with the
extinction-corrected total $V$-band magnitude to determine that
$M_{\rm tot} = 2.8 \times 10^6 \Msun$. The IMBH mass upper limit from
Section~\ref{ss:IMBH} therefore corresponds to $\MBH/M_{\rm tot} \lta
0.43$\%. Table~\ref{t:demo} lists the corresponding constraints for
the other globular clusters that have been well-studied dynamically,
as discussed in Section~\ref{s:intro}. We also include the limit
recently reported by Pasquato \etal (2009) based on a study of the
observed mass segregation in the cluster NGC 2298. An IMBH is expected
to reduce the amount of mass-segregation that is otherwise expected in
a well-relaxed cluster. The suggested IMBH in G1 weighs in at
$\MBH/M_{\rm tot} = (0.32 \pm 0.09)$\%. The four other clusters in the
table only provide upper limits in the range $0.1$--$1$\%. Therefore,
the data allow us to probe in a mass range that is of interest in view
of existing theories.

The IMBHs suspected in some globular clusters may have been similar to
the initial seeds that grew to supermassive black holes (SMBHs) in the
centers of galaxies. It is therefore of interest to compare the
results in Table~\ref{t:demo} to what is known about SMBHs. The masses
of SMBHs scale with either the galaxy bulge mass $M_{\rm bul}$, or
alternatively, with the fourth power of the velocity dispersion
$\sigma$. Haering \& Rix (2004) found that $\MBH / M_{\rm bul} \approx
0.14 \pm 0.04$\%. This is below where it has been possible to probe
for most globular clusters. Only for 47 Tuc is there a limit $\MBH /
M_{\rm tot} \lta 0.13$\% (McLaughlin \etal 2006). Tremaine \etal
(2002) discussed the $\MBH$--$\sigma$ relation proposed for SMBHs. In
Table~\ref{t:demo} we list for each cluster the value of $\MBH$
predicted by their relation. All of the available $\MBH$ upper limits
are at least a factor of 2 above these predictions. Therefore, the
data are insufficient to confirm or rule out the hypothesis that
globular clusters generally have IMBHs that follow the same relations
as established for SMBHs. However, for the case of G1 the suggested
IMBH mass does seem reasonably consistent with these relations
(Gebhardt \etal 2005).

\section{Discussion and Conclusions}
\label{s:conc}

We have presented a detailed dynamical analysis of the star-count,
surface-brightness, and kinematical data available for {\omegaCen},
with a particular focus on the new HST data presented in Paper~I.
Based on the observed profile of the projected density, our models use
the Jeans equation to yield predictions for the projected profiles of
RMS line-of-sight velocity ${\bar \sigma}$ or proper motion ${\bar
\Sigma}$ as function of projected distance $R$ from the cluster
center, in each of the three orthogonal coordinate directions (line of
sight, proper motion radial, and proper motion tangential). We do not
include the effect of mass segregation on the model predictions at
large radii, but we do take into account the observed (partial) energy
equipartition between stars of different masses. The predictions are
compared in a $\chi^2$ sense to observations, to infer the best-fit
values of the model parameters and their error bars.

The model parameters are uniquely determined by the data. The profile
$\beta(r)$ of the intrinsic velocity anisotropy is tightly constrained
by the observed profile $[{\Spmt}/{\Spmr}](R)$. The models are
therefore unaffected by the mass-anisotropy degeneracy that often
plagues line-of-sight velocity studies. The mass-to-light ratio
$\Upsilon$ of the stellar population is fixed by normalization of the
observed $\slos(R)$ and $\Spm(R)$ at large radii. The distance $D$ is
set by comparison of the observed $\Spm(R)$ in mas/yr to the predicted
$\spm(r)$ in km/s. And the presence and properties of any dark mass in
the center, such as an IMBH or dark cluster, are constrained by the
behaviors of $\slos(R)$ and $\Spm(R)$ at small radii.

The new HST proper motion data provide a significant improvement in
our ability to constrain the dynamical models of {\omegaCen}, as
compared to previous modeling efforts (e.g., vdV06; NGB08). The HST
sample increases the number of stars with one or more accurately
measured velocity components by about an order of magnitude. This
allows for the first time the measurement of velocity dispersions at
$R \lta 10''$. We have derived kinematical profiles down to $R \approx
1''$, at which point there cease to be enough stars with measured
proper motions to accurately measure a dispersion. The profile of
$\Spm(R)$ is consistent with being flat in the central $R \lta
15''$. The absence of an increase in RMS stellar velocities towards
the center provides strong constraints on the possible presence of any
dark mass.

The HST star count data from Paper~I provide an upper limit $\gamma
\lta 0.07$ to the central logarithmic cusp of slope of the projected density 
profile. The innermost data points are well matched by previously
published King and Wilson models with a flat core. A generalized nuker
profile fits best at all cluster radii when a shallow cusp is invoked,
but this overpredicts the innermost datapoints at $R \lta
4''$. Dynamical model predictions depend sensitively on whether or not
there is a density cusp, so in our analysis we considered separately
models with a core ($\gamma = 0$) and models with a cusp ($\gamma =
0.05$).

The inferred intrinsic anisotropy profile, distance, and stellar
mass-to-light ratio are all insensitive to the exact light and mass
distributions near the center. We find that the anisotropy changes
from slightly radial near the center to significantly tangential in
the outer parts. The presence of velocity anisotropy is not
surprising, given the long two-body relaxation time of {\omegaCen} and
the existence of other evidence that the cluster is not yet fully
relaxed. The inferred distance is $4.73 \pm 0.09 \kpc$ and the
mass-to-light ratio is $\Upsilon_V = 2.62 \pm 0.06$. The error bars
take into account both the uncertainties in the kinematical data and
the uncertainty in the cusp slope $\gamma$. These results match
extremely well with those obtained by vdV06 using more sophisticated
modeling tools. This agreement supports our assertion that spherical
models based on the Jeans equation are sufficient for the goals of the
present paper, and in particular for constraining the mass
distribution near the center. The facts that the ellipticity and
rotation of {\omegaCen} are both known to decrease towards the center
provide further motivation for this assertion.

Models with a core provide a good fit to the kinematical data without
any dark mass. Since a core is also consistent with the observed
density profile, this implies that the presence of an IMBH is not
required in {\omegaCen}. By contrast, models with a shallow cusp
provide a good fit only when an IMBH is invoked, with $\MBH = (8.7 \pm
2.9) \times 10^3 \Msun$. Indeed, an IMBH can induce a shallow density
cusp (Baumgardt, Makino \& Hut 2005). Without a dark mass, the
predicted RMS velocities in the cusp model decrease towards the
cluster center and the predicted velocity distributions become highly
non-Gaussian. Neither prediction is consistent with the observations.

Upon combination of the predictions for the core and cusp models we
obtain as final result an upper limit to the mass of any possible IMBH
of $1.2 \times 10^4 \Msun$ at $1\sigma$ confidence (or $1.8 \times
10^4 \Msun$ at $3\sigma$ confidence). The $1\sigma$ limit corresponds
to $\MBH/M_{\rm tot} \lta 0.43$\%. This is of similar magnitude as the
range that has now been probed for several other clusters as
well. However, lower values of $\MBH/M_{\rm tot}$ will need to be
probed to definitively assess the most plausible theories for IMBH
formation in globular clusters, or to assess whether globular clusters
may follow the same black hole demographics correlations as galaxies.

Isotropic models are not accurate for {\omegaCen}, since they do not
reproduce the observed deviations of $[{\Spmt}/{\Spmr}](R)$ from
unity. Use of isotropic models to fit ${\bar \sigma}(R)$ yields
spuriously high IMBH masses. Isotropic models do provide a reasonable
guide for calculation of predicted velocity distribution shapes (as
opposed to RMS values) in the central region. The predicted
Gauss-Hermite moments as function of radius for such models fit the
observations in the central arcmin to within $|\Delta h_i| \lta
0.01$. In models with an IMBH, the wings of the velocity distribution
for an aperture on the very center are predicted to be more extended
than in models without an IMBH. However, for cusp models with
plausible IMBH masses we would not have expected more than a single
star in the data set to have a velocity component in excess of $60
\kms$ ($\sim 3.5$ times the cluster velocity dispersion). No such
stars were observed in our data. Given the small number statistics,
this does not provide meaningful new constraints on the IMBH mass for
this particular cluster.

The dark mass invoked in cusp models may in principle be extended.
The $\chi^2$ of the fit yields a limit $a_{\rm dark} \lta 7'' = 0.16
\pc$ on the extent of any dark cluster at $1\sigma$ confidence. The
corresponding central density is $\gta 500$ times larger than the
central density of luminous matter. This would be indicative of quite
an extreme amount of mass segregation. If {\omegaCen} does have a cusp
and a dark mass, then an IMBH may be easier to explain theoretically
than such a concentrated dark cluster.

The IMBH mass suggested by NGB08 is strongly ruled out in all models.
They found $\MBH = 4.0_{-1.0}^{+0.75} \times 10^4 \Msun$ from
isotropic modeling and $\MBH = (3.0 \pm 1.0) \times 10^4 \Msun$ from
anisotropic modeling. By contrast, we find that any IMBH in excess of
$1.8 \times 10^4 \Msun$ is ruled out at $3\sigma$ or higher
confidence. This difference is not surprising, given that our
observations have not reproduced the arguments on which their IMBH
mass was based (namely, their choice of cluster center, their finding
of an increase in dispersion towards that center, and their finding of
a cusp around that center). If one ignores these fundamental
differences, then maybe our results do not appear so different from
those NGB08. When comparing similar models between the two studies
(i.e., cusp models with and without anisotropy), the differences in
the implied IMBH masses are at the $\sim 2\sigma$ confidence level,
given the uncertainties quoted by NGB08 (with our masses being
lower). However, the more important difference is that we find that
core models with no IMBH are perfectly consistent with the data. By
contrast, NGB08 concluded that models with a core and models without
an IMBH were ruled out with least $3\sigma$ confidence. So while NGB08
argued for a significant IMBH detection, we find by contrast that
there is no need to invoke an IMBH in Omega Cen at all. Our finding
that {\omegaCen} cannot have an IMBH as massive as suggested by NGB08
is consistent with arguments presented by Maccarone \& Servillat
(2008). They found that accretion models with the NGB08 IMBH mass
predict much more X-ray and radio emission than is observed from the
cluster center.

To further strengthen the conclusions from this paper and our
understanding of {\omegaCen}, it will be useful to proceed with the
construction of more sophisticated equilibrium models. Methods for
numerical creation of axisymmetric models through orbit superposition
have been developed by, e.g., vdV06, van den Bosch \etal (2006), and
Chaname \etal (2008).  These models can fit the observed differences
between major and minor axis kinematics (discussed in
Appendix~B). They can also make use of the 47,781 high-quality proper
motions along the major axis at $R \approx 1$--6 arcmin, which were
excluded from consideration here (discussed in Appendix~C). This will
yield constraints on the inclination, rotation properties, and the
presence of any kinematic subcomponents. It should also yield further
improved limits on the possible presence of any dark mass in the
center. Construction of non-equilibrium models for {\omegaCen} would
be useful as well. $N$-body, Monte-Carlo, or Fokker-Planck models
might shed light on the evolutionary status of {\omegaCen}, and on how
it arrived at its present structure. This would also provide a more
self-consistent description of the joint effects of mass segregation
and energy equipartition than we have provided here. Moreover, it
would be able to explicitly account for the Brownian motion of a
possible IMBH.


\acknowledgments 

We are grateful to Ivan King, Dean McLaughlin, Pat Seitzer, and Glenn
van de Ven for useful discussions and/or providing data in electronic
format. Suggestions from the referees helped us improve the
presentation of our results.


\appendix

\section{{\bf A.~Extracting Kinematics from Discrete Data}}
\label{s:kindet}

To extract kinematics from a set of discrete data points, we select
the stars that fall in a given area on the sky (e.g., a circular
aperture, annulus, polar wedge, or square aperture). For these stars
we then select the measurements $w_i \pm \Delta w_i$ in a given
coordinate direction (e.g., line of sight, proper motion radial,
proper motion tangential, proper motion major, or proper motion
minor). The $w_i$ may be either velocities in km/s or proper motions
in mas/yr. The goal is to determine the underlying distribution from
which the measurements are drawn. We assume that this is a Gaussian
with mean $W$ and dispersion $\sigma$, and we set out to determine
$W$ and $\sigma$ with their error bars. If the measurement errors
$\Delta w_i$ are all identical then the problem has straightforward
analytical solutions. However, this is not generally the case, so that
more complicated analysis is required (e.g., Appendix~A of vdV06).

We start with a trial estimate of $\sigma$. Each measurement $w_i$ is
then drawn from a Gaussian with dispersion $d_i = (\Delta w_i^2 +
\sigma^2)^{1/2}$. Hence, the velocity W and its error bar $\Delta W$ follow 
from straightforward weighted averaging
\begin{equation} 
  S_1 \equiv \sum_i 1 / d_i^2 , \qquad 
  S_2 \equiv \sum_i w_i / d_i^2 , \qquad
  W = S_2/S_1 , \qquad
  \Delta W = S_1^{-1/2} .
\end{equation}
We then determine the value of $\sigma$ which maximizes the likelihood
\begin{equation}
  L(\sigma) = \prod_i (2 \pi \sigma^2)^{-1/2}
      \exp \lbrace - (w_i - W)^2/ (2 [\Delta w_i^2 + \sigma^2]) \rbrace ,
\end{equation}
by numerically solving the equation $dL/d\sigma = 0$. Having obtained
$\sigma$, we return to an improved determination of $W$ and we iterate
the procedure until convergence. This provides a fixed point iteration
scheme for determination of the joint maximum likelihood solutions for
$W$ and $\sigma$.
 
As described in Appendix~A of vdV06, the maximum likelihood solution
for $\sigma$ yields a subtly biased estimate of the true
dispersion. vdV06 used an analytical approximation to correct for
this. Here we have used a Monte-Carlo procedure instead. After the
determination of $(W,\sigma)$, we draw pseudo-data sets from the
corresponding Gaussian probability distribution. To this we add random
Gaussian errors drawn from the observational uncertainties. We then
analyze numerous such pseudo-datasets in the same fashion as the real
data. The statistics of the Monte-Carlo results provide both an
estimate of the bias in $\sigma$ (which we use to correct our maximum
likelihood estimate) and of the uncertainty $\Delta \sigma$.

After the calculation of $W$ and $\sigma$ has been completed, the RMS
value ${\bar \sigma}$ and its uncertainty can be obtained from
\begin{equation}
  {\bar \sigma} = (W^2 + \sigma^2)^{1/2} , \qquad
  \Delta {\bar \sigma} = [ (W \> \Delta W)^2 + 
                        (\sigma \> \Delta \sigma)^2 ]^{1/2} / {\bar \sigma} .
\end{equation}
For the HST proper motion data, our analysis procedure removed all
mean motions (see Section~\ref{sss:HSTrot}). Therefore, we know that
the mean of the underlying distribution for any data set extracted
from it should be $W=0$. For this case we determined ${\bar
\sigma}$ more efficiently as the solution of the maximum likelihood equation
$dL/d\sigma = 0$ at fixed $W \equiv 0$.

\section{{\bf B.~Major versus Minor axis}}
\label{s:Appmajmin}

In a spherical system one expects the projected kinematics to be the
same along any axis through the center. One also expects that the
projected kinematics for the proper motion coordinate along any fixed
direction, when averaged over a circle, will be independent of the chosen
direction. These expectations are not realized in practice, and this
provides a kinematical signature of the axisymmetry of
{\omegaCen}. vdV06 discussed this in detail for the ground-based
kinematical data. Here we briefly discuss this issue for the HST
proper motion data.

We have extracted major and minor axis subsamples from our HST proper
motion sample, consisting of the stars lying within $10^{\circ}$ from
either axis. When averaged over all radii $R \leq 71.7''$, we find for
the radial proper motion direction that $\Spmr = 0.758 \pm 0.010
\masyr$ along the major axis, and $\Spmr = 0.739 \pm 0.010 \masyr$
along the minor axis. For the tangential proper motion direction we
find that $\Spmt = 0.691 \pm 0.009 \masyr$ along the major axis, and
$\Spmt = 0.764 \pm 0.010 \masyr$ along the minor axis. In a spherical
system the kinematics would be the same along the two axes. In
reality, $\Spmr$ is $(2.5 \pm 1.8)$\% lower along the minor axis than
along the major axis, while $\Spmt$ is $(10.6 \pm 2.1)$\% higher along
the minor axis than along the major axis. There is no statistically
significant variation with radius in these comparisons (for $R \leq
71.7''$).

An alternative way to look at these results is to consider the fixed
major and minor directions on the sky, and to measure the RMS
projected proper motion components $\Spmmj$ and $\Spmmi$ along these
directions, respectively. For the major axis subsample we find $\Spmmj
/ \Spmmi \approx \Spmr / \Spmt = 1.097 \pm 0.021$. For the minor axis
subsample we find $\Spmmj / \Spmmi \approx \Spmt / \Spmr = 1.039 \pm
0.020$. When averaged over the whole sample (i.e., over all position
angles) we find that $\Spmmj / \Spmmi = 1.048 \pm 0.007$. Therefore,
the RMS projected velocity along the major axis is always larger than
that along the minor axis. This was also found by vdV06, who obtained
for the 718 stars with all three velocity components measured that
$\Spmmj / \Spmmi = 1.055 \pm 0.053$ (see their Appendix C and also
their Figure~C.1).

These results are relatively easily understood heuristically. For an
axisymmetric system that is not too far from edge-on, $\Spmmj$ is
primarily related to the pressure along the equatorial plane and
$\Spmmi$ is primarily related to the pressure along the symmetry
axis. The former must exceed the latter in an oblate system, as
dictated by the tensor virial theorem (e.g., Binney \& Tremaine 1987).


\newcommand{\figcapouter}{ Observed major axis profiles of the RMS
projected proper motions $\Spmr$ and $\Spmt$ in mas/yr, as function of
projected radius $R$ from the cluster center. Symbols are the same as
in Figure~\ref{f:kin}b. For the HST data points we included stars
within $10^{\circ}$ from the major axis. For the vdV06 data points
(magenta in the on-line version, with horizontal error bars) we
included stars within a slightly larger angle from the major axis
($17.5^{\circ}$), to increase the sample size. The vdV06 motions were
multiplied by a factor 1.023 to correct to the same characteristic
main-sequence stellar mass as for the HST sample (see
Section~\ref{sss:massseg}). HST data points at $R < 110''$ (closed
dots) are based on the data from the central field, and those at $R >
110''$ (open dots) are based on the major axis field. The figure
focuses on the central region and uses a linear scale in $R$, by
contrast to Figure~\ref{f:kin}b. The curves are predictions based on
the best-fitting anisotropic model with a core and no dark mass, as
discussed in the text. The data for the major axis field are
consistent with expectation, given the detailed modeling of the
central field and the inability of the HST data to measure absolute
rotation in the pmt direction.\label{f:outer}}

\begin{figure*}[t]
\epsfxsize=0.8\hsize
\centerline{\epsfbox{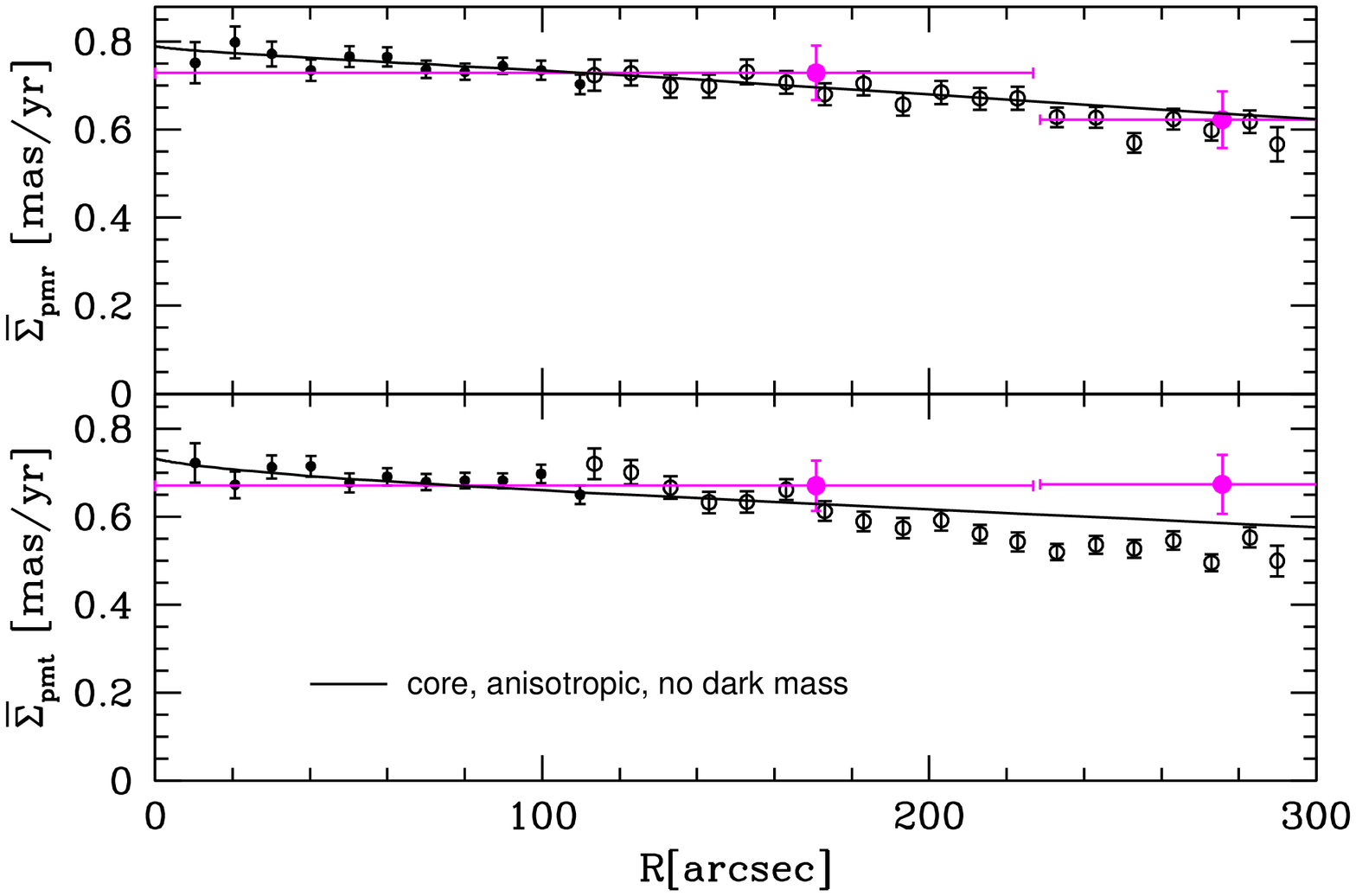}}
\figcaption{\figcapouter}
\end{figure*}


\section{{\bf C.~Proper Motion Kinematics for the Major Axis HST Field}}
\label{s:AppOuter}

As discussed in Section~\ref{sss:HSTsample}, there is also proper
motion data available from Paper~I for another field that was observed
at multiple epochs with HST. This field lies roughly along the major
axis, and therefore does not provide coverage of the full range of
position angles in {\omegaCen}. We therefore did not use it for our
model fitting. However, we address here the major axis profiles that
can be extracted from these data. This allows a comparison with the
central field and our model predictions to check for consistency.

We defined both for the central field and the major axis field a
major-axis subsample consisting of stars lying within $10^{\circ}$
from the major axis. This is the largest wedge for which coverage over
the full range of allowed position angles is guaranteed at the
majority of radii $R$. For the central field we have complete coverage
for $R \leq 115''$. For the major axis field we have complete coverage
for $108'' \leq R \leq 292''$. We extracted the profiles of $\Spmr(R)$
and $\Spmt(R)$ for both subsamples, following the procedures described
in Section~\ref{sss:HSTRMS}. The results are shown in
Figure~\ref{f:outer}. The HST profiles are continuous at the boundary
between the central and major axis fields, as they should be.

At radii $R \leq 71.7''$, we have complete coverage in the central
field over all position angles. We can therefore calculate the ratios
of the projected RMS proper motions measured within $10^{\circ}$ from
the major axis and averaged over an entire circle, respectively. This
yields $(\Spmr)_{\rm maj}/(\Spmr)_{\rm circ} = 1.011 \pm 0.014$ and
$(\Spmt)_{\rm maj}/(\Spmt)_{\rm circ} = 0.938 \pm 0.014$. If we assume
as a simple approximation that these ratios are independent of radius,
then this allows us to compare the predictions of our dynamical models
to the observed major axis profiles. The curves in
Figure~\ref{f:outer} are the model predictions for our best-fitting
anisotropic model with a core and no dark mass (see
Figure~\ref{f:rmsfit}a), scaled by the previously quoted ratios
$(\Spmr)_{\rm maj}/(\Spmr)_{\rm circ}$ and $(\Spmt)_{\rm
maj}/(\Spmt)_{\rm circ}$.

The model curves thus obtained provide an excellent fit to the
$\Spmr(R)$ data for the major axis field. Also, the $\Spmr(R)$ data
for the major axis field are consistent with the data from vdV06,
which are also shown in the figure. By contrast, the fit to the
$\Spmt(R)$ data for the major axis field is good only for $R \lta
200''$. At larger radii, the data fall below the model, and they also
fall below the data from vdV06. We attribute this to the inability of
the HST proper motion data to measure absolute rotation ${\cal V}_{\rm
pmt}$. As a result, we expect to progressively underestimate $\Spmt$
with increasing radius, since ${\cal V}_{\rm pmt}/\Sigma_{\rm pmt}$
increases with radius (see Section~\ref{sss:HSTrot}). The fact that
the data-model comparison is more successful for $\Spmr(R)$ is
explained by the fact that pmr is orthogonal to direction of absolute
rotation, and it is therefore insensitive to it.

Taking into consideration the limitations inherent to the data for the
major axis field, and the results presented here, we have no reason to
believe that detailed modeling of the data for the major axis field
would significantly alter the conclusions of our paper.




{}

\end{document}